\documentclass{aastex62}

\accepted{August 30, 2019}
\submitjournal{ApJ}
\shortauthors{Curiel et al.}

\begin{document} 

\title{Sub-stellar companions of the young weak-line TTauri Star DoAr21}

\correspondingauthor{Salvador Curiel}
\email{scuriel@astro.unam.mx}


\author[0000-0002-0786-7307]{Salvador Curiel}
\affil{
Instituto de Astronom{\'\i}a,
Universidad Nacional Aut\'onoma de M\'exico (UNAM),
Apdo Postal 70-264,
M\'exico, D.F., M\'exico.
}

\author{Gisela N. Ortiz-Le\'on}
\affiliation{Max Planck Institut f\"ur Radioastronomie,
Auf dem H\"ugel 69, D-53121 Bonn, Germany}

\author{Amy J. Mioduszewski}
\affiliation{National Radio Astronomy Observatory,
Domenici Science Operations Center,
1003 Lopezville Road, Socorro,
NM 87801, USA }

\author{Rosa M. Torres}
\affiliation{Centro Universitario de Tonal\'a, Universidad de Guadalajara,
Avenida Nuevo Perif\'erico No. 555, Ejido San Jos\'e Tatepozco, 
C.P. 48525, Tonal\'a, Jalisco, 
M\'exico }
 
\begin{abstract}

The compact, non-thermal emission in DoAr21 has been studied with the VLBA to investigate the possibility that the residuals of the astrometry fitting are due to the reflex motion induced by a possible companion. We find that the fitting of VLBA astrometric observations of DoAr21 improves significantly by adding the orbital motions of three companions. We obtain an improved distance to the source of $134.6\pm1.0$ pc, and estimate that the central star, DoAr21, has a mass of about $2.04\pm0.70~M_{\odot}$. We suggest that DoAr21 represents a unique case where two sub-stellar companions, DoAr21b and DoAr21c ($m_{b}\sim 35.6\pm27.2~M_{jup}$ and $m_{c}\sim44.0\pm13.6~M_{jup}$, respectively), have been found associated to a relatively low mass, pre-main sequence star. In addition, we find that this WTTau star is an astrometric double system, having a low-mass star companion, DoAr21$B$ ($m_{B}\sim0.35\pm0.12~M_{\odot}$), in a relatively eccentric orbit. The orbit of this low-mass stellar companion is compact, while the Brown Dwarfs are located in external orbits. DoAr21$c$ has the strongest astrometric signature in the periodogram, while DoAr21$B$ has a weak but significant signature.  On the other hand, the astrometric signature of DoAr21$b$ does not appear in the periodogram, however, this Brown Dwarf was directly detected in some of the VLBA observations. The estimated orbital periods of DoAr21$B$, DoAr21$b$ and DoAr21$c$ are $P_{B}\sim92.92\pm0.02$, $P_{b}\sim450.9\pm3.8$ and $P_{c}\sim1013.5\pm25.3$ days, respectively. Since the estimated age of this young star is about 0.4$-$0.8 Myrs, the detected Brown Dwarf companion is among the youngest companions observed to date.
 
 \end{abstract}


%

\section{Introduction} \label{sec:intro}

The search for extrasolar  planets has evolve during the past two decades. Several observational technics have proven to be very useful in the search for extrasolar planets, including radial velocity, transits, gravitational microlensing, direct imaging, and even pulsar timing (e.g., \cite{wolszczan92}, \cite{mayor95}, \cite{charbonneau00}, \cite{bond04}, \cite{kalas08}). These various technics are sensitive in different ranges of the orbital period and exoplanet masses, as well as the stelar brighness. In particular, radial velocity and transit searches of low-mass, pre-main seques stars (TTauri Stars) are quite challenging due to the faintness of these objects, their broad (molecular) spectral features, and their ubiquitous variability; indeed, very few exoplanets have been discovered around this kind of stars (e.g.,  \cite{kraus12},  \cite{sallum15}, \cite{donati16}), \cite{yu17}.
A putative Brown Dwarf was recently found to be orbiting a TTauri star \citep{ginski18}, however its orbit lyes outside the circumbinary disk, with a projected orbit of about 210 $AU$. The estimated mass of this companion is quite uncertain, and it could be a very low-mass star.

Astrometry is an additional technique which relies on the positional shift of the star around the center of mass of the orbit due to the gravitational pull of a companion (reflex motion). At present, no firm exoplanet detections have been obtained with this technique. There is only one candidate exoplanet detected with the optical astrometry technique,
however its mass is not well established since the source is a binary system and its mass depends to which star it is associated (\cite{muterspaugh10}).
However, GAIA's astrometric observations have the potential to detect, in the near future, many (probably thousands) exoplanets and Brown Dwarfs associated to solar type and low-mass stars (e.g., \cite{casertano08}, \cite{sozzetti14}, \cite{perryman14}). Astrometric observations can also be carried out in the optical wavelength range with 10 m class telescopes, but require conversion of relative to absolute parallax 
(e.g., \cite{sahlmann16}).

Non-thermal radio continuum emission found associated to TTauri stars and low mass stars (e.g., \cite{phillips91}, \cite{berger01}) opens the possibility to search for sub-stellar companions to this kind of sources.
Furthermore, astrometric searches with the required sensitivity to detect sub-stellar companions can be carried out at radio frequencies using very long baseline interferometry (VLBI); 
see \cite{bower09}, \cite{forbrich09}, \cite{forbrich13}, \cite{gawronski17}.
The principal advantage of VLBI observations would be to obtain absolute accurate stellar positions tied to an extragalactic reference frame, which is crucial for accurately measuring reflex motions, as well as parallaxes and proper motions. 
In this context, the discovery of radio emission from late M and L dwarfs 
(\cite{berger01}; \cite{berger06})
provides a unique opportunity to uncover exoplanets or Brown Dwarf companions to low mass stars and brown dwarfs with M $\leq$0.2 M$_{\odot}$. 
At present, low mass Brown dwarfs (several tens of Jupiter masses) have been found orbiting a few ultracool stars 
(e.g., \cite{sahlmann13}),
 however, sub-stellar companions have not been found orbiting TTauri stars.

 DoAr21 is an unusual X-ray bright ($\sim$$10^{32}$ erg/s) weak-line TTauri star (WTTS;  \citet{neuhaeuser95}) in the Ophiuchus Molecular Cloud with a spectral type between K0-K2, an age of ~ 0.8 Myrs, a mass of $\sim$$2.2 \,M\odot$ (however, it could also be consistent with a binary system with masses of  $\sim$$1.8 \,M\odot$ and an age of ~ 0.4 Myrs) and fairly obscured (A$_{V}$ $\sim$6-7) (\citet{jensen09}, and references therein). It also exhibits modest 24-micron excess, suggestive of a circumstellar disk, which is unusual for WTTS, while showing no indication of near-IR excess at shorter wavelengths. An extended asymmetric ring structure is detected at NIR H2 emission \citep{panic09}, which is at $\sim$73-219 AU away from the central star. This is consistent with the estimation of a cavity size of $\sim$70 AU \citep{vdmarel16}.  \citet{jensen09} found that the PAH, H2, and mid-infrared excess are most likely associated with a small-scale photo-dissociation region (PDF) located $\sim$100$'$s of AU from the star that is perhaps excited by the UV emission of DoAr21.  \citet{james16} found that the SED of this source is consistent with a $3.5 \,M\odot$ star and a very low mass disk of  $\sim$$4.5\times10^{-5} \,M\odot$ with an inclination of about 81 deg.

DoAr21 was first detected at radio wavelengths with the VLA by \cite{feigelson85}, who found  that the radio flux density of this source had a steep increase in time scales of a few hours. Later on DoAr21 was detected with the VLBA by \cite{phillips91}, showing similar variability in its radio flux density and indicating a non-thermal origin.
This source displays an X-ray spectrum typical of pre-MS stars in Orion \citep{preibisch05} that is characterized by components with temperatures of $\sim$1 and 3 keV. Flares were detected during two independent $\sim$100 ks observations of DoAr21. One epoch displayed small-scale flares superimposed on a slowing declining light curve \citep{gagne04} whereas the more recent observations displayed an impulsive flare during which the temperature of the X-ray emitting plasma increased by a factor of $\sim$2 \citep{jensen09}. Indeed, \citet{jensen09} concluded that DoAr 21 exhibits X-ray flares at a rate of nearly one per day. This high rate may reflect the increased likelihood of flaring in small-separation TTS binary systems, wherein the coronal activity of each component may be influenced by the interactions of the stars$'$ magnetospheres \citep{stelzer00}. 
It is suggestive that there also seems to be a correlation between excess of non-thermal radio emission in young stars and close binarity, and in fact, DoAr21 is one of two stars which are always significantly brighter than all other YSOs in the Ophiuchus star formation complex in non-thermal radio emission, the other star, S1, is a known binary \citep{ortizleon17}.

The compact, non-thermal emission in DoAr21 has been studied at radio wavelengths with the VLBA (\citet{loinard08}; \citet{ortizleon17}; \citet{ortizleon18}), which provided an accurate estimate of the distance for this radio continuum source of 135.4 +/- 4.3 pc. This error is 2-3 times larger than typical errors obtained for the other sources in \citet{ortizleon17}, which suggests that something is going on with this source. One of the VLBA observations reported by \citet{loinard08}  shows that DoAr21 could be a binary, with a projected separation of about 5 mas ($\sim$0.6 AU). The orbital parameters are uncertain, but Loinard et al. estimated that the semi-major axis could be of the order of 1-2 AU. Since the binary pair has been in principle detected at only one epoch thus far, there is no information about the mass ratio or the optical luminosity ratio. Ascribing all of the luminosity to the primary star yields a mass of $\sim$$2.2 \,M_\odot$ and an age of $\sim$0.4 Myrs, and if the luminosity is split equally between the two stars, the mass of each star is $\sim$$1.8 \,M_\odot$ and an age of $\sim$0.8 Myrs \citep{jensen09}.

In this paper we investigate the possibility that the reflex motion due to a companion orbiting DoAr21 is the responsible of the large residuals of the astrometric fitting of the multi-epoch non-thermal radio emission. In this paper we present new VLBA observations of this source taken over a period of 9 months. We have also re-calibrated previous VLBA observations of this source to search for evidence of the putative companion of this WTTS. The new and archival observations are presented in Section 2, and the new detections and the reanalyzed data are also described in section 2. In Section 3 we present the algorithms that we use in the search for companions of this source. The results and the discussion are presented in Sections 4 and 5, and we present our main conclusions in Section 6.


\section{Observations}

DoAr21 has been observed with the Very Long Baseline Array (VLBA) during several multi-epoch campaigns. We found in the NRAO archive public data from projects BL128 and BT093. The observations of project BL128  were taken between September 2005 and December 2006, while project  BT093 observed between July and September  of 2007.   Both projects were taken at 8.4~GHz with 32~MHz of bandwidth and observed  the quasar J1625$-$2527 as the phase reference calibrator.  A subset of these observations were published in \cite{loinard08}. In total, they covered 17 epochs, from which we use only the 3 epochs that clearly detected the two sources. We also use the  positions of DoAr21 published in \cite{ortizleon17}, which correspond to epochs observed from  August 2012 to September 2017 under program BL175 (GOBELINS project). Finally, four new observations were obtained under program BC237 on May, August  and November 2018, and February 2019. The new data were taken at 8.4~GHz with 256~MHz of bandwidth and, similar to BL175, used J1627--2427 for phase reference calibration, which is located in projection towards the Ophiuchus core  \citep{dzib13}.

Project BC237 included two observing blocks of about 30 minutes each spent on calibrators distributed over a wide range of elevations. These scans (the so-called geodetic-like blocks) were used to estimate the multi-band delays, i.e.\ the phase slope with frequency, which are introduced by tropospheric and clock errors. 

We conduct an homogeneous calibration for the BC237 data  and the archival data (BL128 and BT093) using AIPS \citep{greisen03}. This means that the same calibration steps are applied to each epoch as follows. First, all scans with elevations below $10^{\rm o}$ are flagged.  Ionosphere dispersive delays are  removed using Global Positioning System (GPS) models of the electron content in the ionosphere which are downloaded from the Crustal Dynamics Data Information System (CDDIS) archive. Corrections to the Earth Orientation Parameters and amplitude corrections for digital sampling effects of the correlator are also applied.  Instrumental single-band delays are then determined and removed using fringes detected on a single scan on the calibrator J1625$-$2527. The correction for the  bandpass shape is then  done using the same calibrator. 
To  finish the amplitude calibration, we use the provided gain curves and system temperature tables to derive the System Equivalent Flux Density (SEFD) of each antenna.  After applying corrections for the rotation of the RCP and LCP feeds, the multi-band delays are derived by fringe-fitting the scans of the geodetic-like blocks and used to estimate the tropospheric and clock errors, which are removed from the data.

To finish the phase calibration, fringe fitting is run on the  phase reference calibrator to find residual phase rates.  In order to take the structure of the reference calibrator into account, this last step is repeated using a self-calibrated image of the calibrator as a source model. Finally, the calibration tables are applied to the data and images of the target are produced using a pixel size of $50-100~\mu$as and pure natural weighting. 

Table \ref{tab:jmfit} lists the positions of DoAr~21 taken from \cite{ortizleon17}. Source positions at the other epochs were obtained by fitting a Gaussian model to the source brightness distribution in the produced images using the AIPS task JMFIT. The angular resolution of the final images of BC237 was typically $\sim$2~mas and the noise level was $\sim25~\mu$Jy~beam$^{-1}$ in the best case.

\begin{figure*}
   \centering
   \plotone{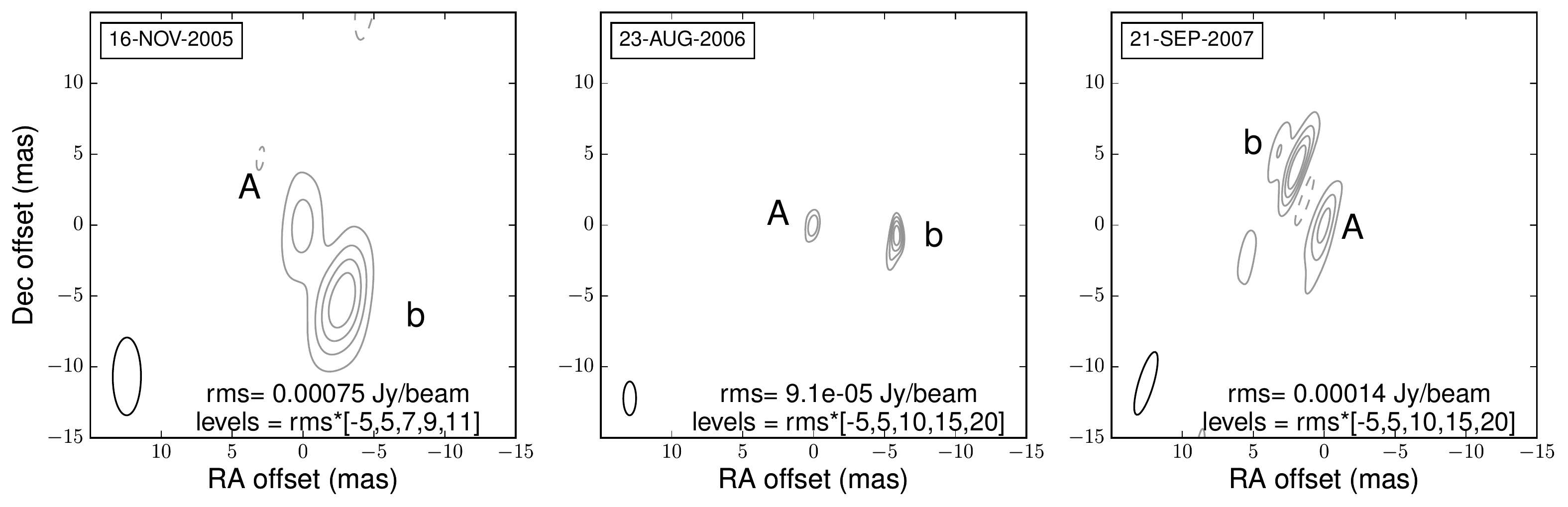}
    \caption{DoAr21 contour maps of the three epochs when two sources were detected. Label A
     indicates the position of the primary source (host star DoAr21) and label b indicates de position of the secondary source (DoAr21$b$).}
              \label{fig_1}%
    \end{figure*}

We detect two sources on 2005 November 16, 2006 August 24 and 2007 September 21, while a single source is seen in the remaining epochs of projects BL128 and BT093, and the same for BC237. Counter plot maps of the three epochs are presented in Figure~\ref{fig_1}.

It is important to mention that the observed epochs in project BL175 were calibrated following the same procedure that we use here (see  \cite{ortizleon17} for more details). Thus, it is expected that the positions of the source detected have a similar precession as those obtained with the new observations of this source and that we report here. In addition, the calibration procedure used for the data observed from projects BL128 and BT093 was different than the one we used for the new observations. However, in this study we only use the relative position of the two sources that we have detected in  three epochs.


\section{Fitting of the astrometric data}

\subsection{Least-squares Periodograms}

The most popular method to search for periodicities in data is the so-called Lomb-Scargle periodogram. However, this method performs optimally only under an important implicit assumption: all the other signals (e.g., linear trend, an average offset, etc.) can be subtracted from the data without affecting the significance of the signal. This assumption does not hold for astrometry because the proper motion and the parallax are also a significant part of the signal and they typically correlate with the periodic motion of a companion (see \citet{black82}).

Following the procedure presented by \citet{anglada10}, we use instead a circular least-squares periodogram (CLS periodogram). In this approach, the weighted least-squares solution is obtained by fitting all the free parameters in the model for a given period. The sum of the weighted residuals divided by N is the so-called $\chi^{2}$ statistic, where N is the number of data points. Then, each $\chi_{P}^{2}$ of a given model with $k_{P}$ parameters can be compared to the $\chi_{0}^{2}$ of the null hypothesis with $k_{0}$ free parameters by computing the power, $z$, as

\begin{equation}
z(P)  =  \frac{(\chi^{2}_{0} - \chi^{2}_{P})/(k_{P} - k_{0})}{\chi^{2}_{P}/(N_{obs} - k_{P})}, 
\end{equation}

\noindent
where a large $z$ is interpreted as a very significant solution. The values of $z$ follow a Fisher $F$-distribution with $k_{P} - k_{0}$ and $N_{obs} - k_{P}$ degrees of freedom
\citep{scargle82, cumming04}. Even if only noise is present, a periodogram will contain several peaks \citep[as an example]{scargle82} whose existence have to be considered in obtaining the probability that a peak in the periodogram has a power higher than $z(P)$ by chance, which is the so-called false alarm probability (FAP):

\begin{equation}
FAP = 1 - (1 - Prob[z > z(P)])^{O},
\end{equation}

\noindent
where $O$ is the number of independent frequencies. In the case of uneven sampling, $O$ can be quite large and is roughly the number of periodogram peaks one could expect from a data set with only Gaussian noise and the same cadence as the real observations. We adopt the recipe $O \sim 2\Delta T/P_{min}$ given in Cumming (2004, Section 2.2), where $\Delta T$ is the time span of the observations and $P_{min}$ is is the minimum period searched. For instance, assuming that $\Delta$$T$ = 2300 days and $P_{min}$ = 20 days, the astrometric data is expected to have  $O \sim 230$  peaks.

In our case, the null hypothesis is the basic kinematic model with $k_{0} = 5$: two reference coordinates, two proper motions and the parallax. As a first approach, our simplest non-null hypothesis considers circular orbits. For a given period, the number of free parameters is then $k_{P} = 9$: five kinematic parameters plus the four Thiele Innes elements $X1$, $Y1$, $X2$  and  $Y2$ \citep{green93}. 

As a second approach, we do a least-squares fit of a full Keplerian orbit to the astrometric data, and if we obtain a reasonable fit for the eccentricity of the orbit, we recalculate the periodogram with the eccentricity fixed to this value. To find the new least-squares periodogram, we find the time passage at the periastron $\tau$ using a Monte Carlo method for each given period (and the eccentricity fixed), and we perform a least-squares periodogram sampling of a grid of fixed eccentricity-period ($eP$; where $e$ is fixed either to 0 or the fitted value) pairs and fitting all other parameters. For each $eP$ pair and estimated $\tau$, $k_{P}$ is 9: the null-hypothesis parameters plus all the other Keplerian elements (using the Thiele Innes elements): $\omega$, $\Omega$, $a_{1}$, and $i$, where $a_{1}$ is the semi-mayor axis of the orbit of the star due to the companion. As mentioned in section 2, in the fitting of the data,  we only use the  positions of DoAr21 published in \cite{ortizleon17} and those that we report here, which cover a time span of about 2375 days. Since the model is linear in all 9 parameters, the power can be efficiently computed for many periods between 20 days and 3500 days (i.e., a period larger than the time span of the observations) to obtain the familiar representation of the periodogram (see Figure~\ref{fig_2}).  

\subsection{Astrometric Fits}

Here, we present a model of $\mu$as astrometric data of the sort that can be provided by Very Long Baseline Interferometry (VLBI), such as the VLBA, as well as  GAIA (and in the future by the next generation Very Large Array (ngVLA)).

The source barycentric two-dimensional position is described as function of time, accounting for the (secular) effects of proper motions ($\mu_\alpha$ and $\mu_\delta$), the (periodic) effect of the parallax $\Pi$, and the (Keplerian) gravitational perturbation induced on the host star by one or more companions (low mass stars, sub-stellar companions, or planets; mutual interactions between companions are not taken into account). The proper motions and the annual parallax terms can be expressed as follows:

\begin{eqnarray}
\alpha(t) & = & \alpha_{0} + (\mu_\alpha cos(\delta)) (t - t_{0}) + 0.5 (a_\alpha cos(\delta))(t - t_{0})^2
+ \Pi F_\alpha(t) + \sum G_\alpha(t), \\
\delta(t) & = &  \delta_{0} + \mu_\delta (t - t_{0}) + 0.5 a_\delta (t - t_{0})^2 + \Pi F_\delta(t)
+ \sum G_\delta(t),
\end{eqnarray}
 
 \noindent
where ($\alpha_{0}$, $\delta_{0}$) is a reference position, $t_{0}$ is a reference time (usually the mean epoch of the multi-epoch observations), and $a_\alpha$ and $a_\delta$ are acceleration terms needed to take into account the dynamical effect induced by long-period stellar companions to the primary (e.g., when the primary is part of an unseen long-period stellar binary; these acceleration terms can also be necessary for close-by high-velocity stars). However, the acceleration terms are not always necessary. $F_\alpha$ and $F_\delta$ refer to the astrometric displacement due to the parallax in $\alpha$ and $\delta$ directions during observations, which are given by  \citep{seidelman92}:

\begin{eqnarray}
F_\alpha (t) & = & (X sin(\alpha) - Y cos(\alpha))/(15 cos(\delta)), \\
F_\delta (t)  & = & X cos(\alpha) sin(\delta) + Y sin(\alpha) sin(\delta) - Z cos(\delta),
\end{eqnarray}
  
\noindent
here ($X,Y,Z$) represent the Cartesian components in equatorial coordinates of the position of the observatory at  the time of the observations, t, with respect to the solar system barycenter (in units of AU when $\Pi$ is in arcsec) and $\alpha$ and $\delta$ are the coordinates of the barycentric place of the source at each epoch. These values for the Earth are available from the NASA Jet Propulsion Laboratory Solar System ephimerides$\footnote{http://ssd.jpl.nasa.gov}$.
  
The term ($G_\alpha(t)$,$G_\delta(t)$) is the induced Keplerian orbit due to an unseen companion, and the parallax factors are defined using the classic formulation by \citet{green93}.
The Keplerian orbit of each companion is scaled and projected onto the plane of the sky through (\citet{green93}):

\begin{eqnarray}
G_\alpha(t) & = & r [cos(\nu + \omega) sin(\Omega) + sin(\nu + \omega) cos(\Omega) cos(i)], \\
G_\delta(t) & = & r [cos(\nu + \omega) cos(\Omega) - sin(\nu + \omega) sin(\Omega) cos(i)],
\end{eqnarray}

\noindent
where $i$ is the inclination of the orbital plane (such that $i$ = 0 corresponds to a face-on, anti-clockwise orbit), $\omega$ is the longitude of the periastron, $\Omega$ is the position angle of the line of nodes, $\nu$ is the true anomaly, and $r$ is the radius vector, which can be expressed in terms of the true anomaly $\nu$, or the eccentric anomaly $E$, using the dynamical equations

\begin{equation}
r  =  \frac{a (1 - e^{2})}{(1 + e cos(\nu))} 
       =    a (1 - e cos(E)),
\end{equation}

\noindent
where $e$ is the eccentricity, and $a$ is the apparent semi-major axis of the star's orbit around the systemic barycenter, i.e., the astrometric signature.
The eccentric anomaly is the solution of Kepler's equation:

\begin{equation}
E - e sin(E) = M,
\end{equation}

\noindent
with the mean anomaly $M$, expressed in terms of the orbital period $P$ and the epoch of the periastron passage $\tau$:

\begin{equation}
M = \frac{2 \pi (t - \tau)}{P}.
\end{equation}

\noindent
Finally, the true anomaly $\nu$ is function of the eccentricity and the eccentric anomaly:

\begin{equation}
tan\left(\frac{\nu}{2}\right) = \left(\frac{1 + e}{1 - e}\right)^{1/2} tan\left(\frac{E}{2}\right).
\end{equation}

Note that since $a$ is the apparent semi-major axis of the star's orbit in units of arcsec, its relationship to the mass of the secondary depends on the astrometric method used. For astrometric perturbations due to an unseen companion, we have from Kepler's Third Law:

\begin{equation}
a^{3} = \frac{\Pi^{3} m_{2}^{3}}{(m_{1} + m_{2})^{2}} P^{2},
\end{equation}

\noindent
where $a$ is measured in arc-seconds when $P$ is measured in years, $\Pi$ is the parallax in arc-seconds and $m_{2}$ is the mass of the unseen companion and $m_{1}$ is the mass of the star, both in solar masses.

The unknown parameters are a reference position, proper motions, acceleration terms, parallax, and $7\times$$n_{p}$ orbital elements, where $n_{p}$ is the number of companions that are fitted ($P$, $\tau$, $e$, $\omega$, $\Omega$, $a$ and $i$ for each companion).  A total of 14 free parameters in the case of a single companion, 21 free parameters for two companions, etc.

In the case of an astrometric binary (when the two stars in the binary are detected), it is necessary to fit the Keplerian parameters of the binary, the proper motions of the system's barycenter and the parallax simultaneously (see above).  For the secondary, $\omega$ is rotated 180 deg, and $a_{2}$ is used instead, which is scaled from $a_{1}$ by the mass ratio $q$=($m_{2}/m_{1}$).
In this case, two additional unknown parameters, $a_2$ and $q$, need to be taken into account to fit the secondary component, making, in this case, a total of 16 free parameters. However, in most cases only 14 free parameters are needed. The acceleration terms are only needed when there is something that perturbes the motion of this compact system, which could be the case when the binary is part of a wider multiple system.

In the case of having relative astrometry observations of the binary plus absolute observations of only one of the components in the binary (we will call it the main component), the fitting procedure is similar to the case of a binary system, but in this case the two additional parameters ($\omega_{2}$ and $a_{2}$ or $q=(m_{2}/m_{1})$) are used to fit the relative astrometry, the other orbital parameters are the same used for the absolute astrometry: $P$, $\tau$, $e$, $\omega$, $\Omega$, $a$ and $i$. In this case there are also 16 free parameters. In the case that the primary is the most massive component in the system ($q << 1$), 7 additional parameters can be added to fit a second companion, making a total of 23 free parameters. Since the acceleration terms are only needed when the compact system is part of a wider multiple system (e.g., a triple or quadruple system, where the other stellar components are farther away from the observed compact system and have orbital periods much more larger than the orbital period in the compact system), in general only 14 free parameters (in the case a single companion) or 21 free parameters (in the case of two companions) will be fitted.  

\subsection{Least-squares Fitting Algorithm}

In this case, we follow the procedure described in the previos section, but here the orbital elements ($\omega$, $\Omega$, $a_{1}$ and $i$) are obtained using the Thiele Innes elements $X1$, $Y1$, $X2$, and $Y2$ \citep{green93} instead of equations 7 and 8. In this case:

\begin{eqnarray}
G_\alpha(t) & = & X1  x(t) + X2  y(t), \\
G_\delta(t)  & = & Y1  x(t) + Y2  y(t),
\end{eqnarray}

\noindent
where the Thiele-Innes constants  are expressed as:

\begin{eqnarray}
X1 & = &  a[cos(\omega) sin(\Omega) + sin(\omega) cos(\Omega) cos(i)], \\
Y1 & = &  a[cos(\omega) cos(\Omega) - sin(\omega) sin(\Omega) cos(i)], \\
X2 & = & a [-sin(\omega) sin(\Omega) + cos(\omega) cos(\Omega) cos(i)], \\
Y2  & = &  a[-sin(\omega) cos(\Omega) - cos(\omega) sin(\Omega) cos(i)].
\end{eqnarray}

The elliptical rectangular coordinates $x(t)$ and $y(t)$ are given in terms of the dynamical equations (equation 9) and the true anomaly $\nu$ (equation 12) by

\begin{eqnarray}
x(t) & = & \left(\frac{r}{a}\right) cos(\nu), \\
y(t) & = & \left(\frac{r}{a}\right) sin(\nu).
\end{eqnarray}
 
The linear parameters (reference position, proper motions, acceleration terms, parallax, and $4\times$$n_{p}$ orbital elements) are fitted by Matrix Inversion, while the non-linear parameters ($P$, $\tau$ and $e$) are found using a Montecarlo Method. We use the correlation matrix to obtain the uncertainty of the fitted parameters.

\subsection{AGA Fitting Algorithm}

The Asexual Genetic Algorithm (AGA) fitting procedure that we use here is similar to that described by  \citet{canto09}  and  \citet{curiel11}, where the fitting procedure was used to find Keplerian orbits, using the radial velocities (RV) of the host star. This method can be extended to the problem of fitting  the Keplerian elements of an orbit to astrometric data,
such as those that can be obtained with the VLBA, GAIA and, in the  future, with the ngVLA. 
As described by \citet{canto09}, the curve or model fitting is essentially an optimization problem. Given a discrete set of N data points ($\alpha_{i},\delta_{i}$) with associated measurement errors $\sigma_{i}$, one seeks for the best possible model (in other words, the closest fit) for these data using a specific form of the fitting function, ($\alpha(t),\delta(t)$). This function has, in general, several adjustable parameters, whose values are obtained by minimizing a $''$merit function$''$, which measures the agreement between the data ($\alpha_{i},\delta_{i}$) and the model function ($\alpha(t),\delta(t)$). 
The maximum likelihood estimate of the model parameters ($c_{1}, ..., c_{k}$) is obtained by minimizing the $\chi^{2}$ function:

\begin{equation}
\chi^{2}_{min}  = \sum_{i=1}^{N} \left( \frac{\alpha_{i} - \alpha(t_{i}; c_{1}, ..., c_{k})}{\sigma_{i}}\right)^{2} 
              + \sum_{i=1}^{N} \left( \frac{\delta_{i} - \delta(t_{i}; c_{1}, ..., c_{k})}{\sigma_{i}}\right)^{2},   
\end{equation}

\noindent
where each data point ($\alpha_{i},\delta_{i}$) has a measurement error that is independently random and distributed as a nominal distribution about the $''$true$''$ model with standard deviation $\sigma_{i}$.
We then apply the AGA method to search for the best solution using  this initial guess.
Here we have made two important improvements to the original algorithm. 
First, we do an initial search in the hipper$-$cube space of possible solutions to find an initial guess for the parameters that are fitted. 
Second, the error estimates, $\sigma(c_{j})$, for the fitted parameters $c_{j}$ can be estimated as the projection of the confidence region of the m-dimensional space parameter for which $\chi^{2}$ does not exceed the minimum value by an amount $\Delta(m,\alpha)$, where $\alpha$ is the significance level (0 $<$ $\alpha$ $<$ 1). 
Following \citet{avni76} and \citet{wall03}, the probability

\begin{equation}
Prob [\chi^{2} -  \chi^{2}_{min} \le  \Delta(m,\alpha)] = \alpha
\end{equation}

\noindent
is that of a chi$-$square distribution with $m$ degrees of freedom. Thus, $\Delta(m, \alpha)$ is the increment of $\chi^{2}$ such that if the observation is repeated a large number of times, a fraction $\alpha$ of times the values of the parameters fitted will be inside the confidence region, i.e., in the interval $c_{j}$ $\pm$ $\sigma(c_{j})$ (see \citet{estalella17}, and references therein).

In this case, we also follow the procedure described in  previos section 3.2.
This fitting procedure allow us to fit simultaneously all free parameters. However, in the case of a single star, since we do not know a priori if it has a single or multiple companions, we first fit the proper motions and the parallax of the star using the least-squares method and the AGA method, and if necessary we include the acceleration terms. We then analyze the residuals in order to find if there may be an astrometric companion orbiting around the star. To do this, we compute the least-squares periodogram of the astrometric data comparing the null solution and the Keplerian solution (see Section 3.1). If the periodogram shows possible companions (e.g., significant peaks with FAPs smaller that $1\%$), we then include a possible companion in the fitting procedure of the data, using again both fitting methods: least-squares and AGA algorithms. We apply this procedure for each signal present in the periodogram. 
In the case that there is more than one signal in the periodogram, and there are enough observations (enough data points) to fit simultaneously all the required free parameters, 
we fit simultaneously more than one astrometric companion. In the case that more than one component is detected with the VLBA, we fit simultaneously the two components, either by doing full astrometry to the main component, or by combining the relative astrometry of the system and the absolute astrometry of one of main component.

Here we follow the standard nomenclature to name the host star and the companions that were detected: we use a capital letter for the host star and  low-mass stellar companions, and a lower case letter for sub-stellar companions. Thus, in what follows, DoAr21 or DoAr21$A$ is the host star, DoAr21$B$ is the low-mass stellar companion, DoAr21$b$ is the inner sub-stellar companion, and DoAr21$c$ is the outer sub-stellar companion.

\subsection{Applying the least-squares and the AGA algorithms to other data sets}

Before fitting the astrometric data that we present here, we used both algorithms, as well as the least-squares periodogram, to fit the data sets of several sources previously investigated by our team in the Ophiuchus region \citep{ortizleon17}.  We fitted single sources (when only one source was detected with the VLBA) and binary systems (when two sources were detected with the VLBA). We found that our results are in good agreement with those reported by \citet{ortizleon17}.
As an example, we present in Table \ref{tab:binary_fit} the fit of the binary system LFAM~15. The main difference between our fit and that obtained by \citet{ortizleon17} is the time of the periastro of the orbit, which is offset by two orbital periods of the binary system. We obtained a periastro time near the center of the observing interval, while \citet{ortizleon17} obtained a periastro time near the beginning  of the observing interval.


\begin{table*}[h]
\caption{Measured VLBA Source Positions}             
\label{tab:jmfit}
\centering                          
\begin{tabular}{ccccccc}        
\hline\hline                 
Julian date & $\alpha$ (J2000.0) & $\sigma_\alpha$ & $\delta$ (J2000.0) & $\sigma_\delta$ & Reference  \\    
\hline                        
&& Primary &&& \\
\hline                        
2456158.56345 & 16  26  3.00879650 & 0.00000023 & $-$24   23  36.532098 & 0.000008 & 1 \\
2456271.25813 & 16  26  3.00899222 & 0.00000045 & $-$24   23  36.541056 & 0.000014 & 1 \\
2456537.53271 & 16  26  3.00730547  & 0.00000110 & $-$24   23  36.559579 & 0.000037 & 1 \\
2456718.03765 & 16  26  3.00766323 & 0.00000030 & $-$24   23  36.575357 & 0.000010 & 1 \\
2456938.43345 & 16  26  3.00579982 & 0.00000097 & $-$24   23  36.589569 & 0.000036 & 1 \\
2457081.04343 & 16  26  3.00621579 & 0.00000166 & $-$24   23  36.602285 & 0.000058 & 1 \\
2457300.44295 & 16  26  3.00441795 & 0.00000725 & $-$24   23  36.615740 & 0.000215 & 1 \\
2458257.82118 & 16  26  3.00114839 & 0.00000028 & $-$24   23  36.688565 & 0.000009 & 2  \\
2458335.60840 & 16  26  3.00026270 & 0.00000402 & $-$24   23  36.692870 & 0.000088 & 2 \\
2458433.34082 & 16  26  3.00019279 & 0.00000619 & $-$24   23  36.700039 & 0.000131 & 2 \\
2458534.06442 & 16  26  3.00050166 & 0.00000082 & $-$24.  23  36.709262 & 0.000026 & 2 \\
\hline                        
\hline                        
&& Binary System &&& \\
\hline                        
&& Primary &&& \\
\hline                        
2453691.33229 & 16 26 3.01909977 & 0.000005314 & $-$24 23 36.343748 & 0.000153 & 2  \\
2453971.56511 & 16 26 3.01741929 & 0.000004120 & $-$24 23 36.368881 & 0.000124 & 2  \\
2454365.48657 & 16 26 3.01575121 & 0.000002832 & $-$24 23 36.402405 & 0.000107 & 2  \\
\hline                        
&& Secondary && \\
\hline                        
2453691.33229 & 16 26 3.01889886 & 0.00000304 & $-$24   23  36.349153 & 0.000063 & 2  \\
2453971.56511 & 16 26 3.01698794 & 0.00000267 & $-$24   23  36.369931 & 0.000114 & 2  \\
2454365.48657 & 16 26 3.01588946 & 0.00000202 & $-$24   23  36.398070 & 0.000067 & 2 \\
\hline                                   
\end{tabular}
\tablenotetext{}{(1)~\citet{ortizleon17}; 
(2) This work.
}

\end{table*}


\begin{table*}[h]
\caption{Binary Astrometry Fits\tablenotemark{a} }             
\tablenotemark{a}
\label{tab:binary_fit}
\centering                          
\begin{tabular}{lcc}        
\hline\hline                 
  & LFAM15   &    \\    
\hline\hline                 
Parameter & This work  &  \citet{ortizleon17}  \\    
\hline                        
& Parameters Fitted & \\
\hline                        
$\mu_x$ (mas/yr) &   $-$6.303 $\pm$ 0.020 & $-$6.31 $\pm$ 0.02 \\
$\mu_y$ (mas/yr)   &  $-$26.964 $\pm$ 0.050 & $-$26.95 $\pm$ 0.05  \\
$\Pi$ (mas)           & 7.259 $\pm$  0.079     & 7.253 $\pm$ 0.054  \\
$P$ (days)            &  1308.75 $\pm$ 10.39            & 1311.61 $\pm$ 6.68  \\
$T_{0}$ (days)        &   2014.1931 $\pm$  0.076     & 2007.008 $\pm$ 0.039  \\
$e$                       &   0.5292 $\pm$ 0.0075          &  0.528 $\pm$ 0.005  \\ 
$\omega$ (deg)    &  56.09 $\pm$ 0.52               & 55.54 $\pm$ 1.02     \\
$\Omega$ (deg)   &  338.01 $\pm$ 0.40                 & 337.93 $\pm$ 0.81       \\
$a_1$ (mas)         &  7.760 $\pm$ 0.070           &  $...$    \\
$i$ (deg)              &   110.24 $\pm$ 0.27               & 110.30 $\pm$ 0.49           \\
$\omega_2$ (deg)    &   236.09 $\pm$ 0.52               & 235.54 $\pm$ 1.02     \\
$a_{2}$ (mas)         &   8.637 $\pm$ 0.091           &  $...$    \\
$a$ (mas)         &   16.40 $\pm$ 0.16           & 16.40 $\pm$ 0.13     \\
\hline                        
& Other Parameters & \\
\hline                        

$D$ (pc)                   &  137.77 $\pm$ 1.48      & 137.9 $\pm$ 1.0   \\
$m  ~(M_\odot)$  &   0.898 $\pm$ 0.042          & 0.89 $\pm$ 0.01                \\
$m_{1}  ~(M_\odot)$  &   0.473 $\pm$ 0.022  & 0.469 $\pm$ 0.015       \\
$m_{2} ~(M_\odot)$   &   0.425 $\pm$ 0.019 & 0.421 $\pm$ 0.010      \\
$a_{1} ~(AU)$            &   1.069 $\pm$ 0.021 &  $...$  \\
$a_{2} ~(AU)$           &   1.190 $\pm$ 0.025 & $...$   \\

$\chi^2$, $\chi^2_{red}$   &   75.53, 3.43 & $...$, $..$  \\
%
\hline                                   
\end{tabular}
\tablenotetext{a}{ The parameters presented here were obtained with AGA. Very similar results were obtained with the least-squares fitting method. The sub-index 1 corresponds to the main component and the sub-index 2 corresponds to the secondary component.}

\end{table*}


\begin{table*}[h]
\caption{Absolute Astrometry Fits: One companion\tablenotemark{a} }             
\tablenotemark{a}
\label{tab:astrofit}
\centering                          
\begin{tabular}{lccc}        
\hline\hline                 
Parameter & Single Star Solution & DoAr21$c$ & DoAr21$B$   \\    
\hline                        
&& Parameters Fitted & \\
\hline                        
$\mu_x$ (mas/yr) & $-$19.6953 $\pm$ 0.0020  &  $-$19.7089 $\pm$ 0.0028 & $-$19.5565 $\pm$ 0.0028 \\
$\mu_y$ (mas/yr)  & $-$26.9477 $\pm$ 0.0050 &  $-$26.9637 $\pm$ 0.0069 & $-$26.8091 $\pm$ 0.0069  \\
$acc_x$ (mas/yr/yr)  &        $... $                        &            $.....$               & 0.0335 $\pm$ 0.0020  \\ 
$acc_y$ (mas/yr/yr)  &       $.....$                        &            $.....$               & $-$0.0015 $\pm$ 0.0048    \\ 
$\Pi$ (mas)           & 7.5330 $\pm$ 0.0068         & 7.5001 $\pm$  0.0095     & 7.4482 $\pm$ 0.0095  \\
$P$ (days)            & $.....$                                  & 1018.43 $\pm$ 3.98            & 92.892 $\pm$ 0.030  \\
$T_0$ (days)        &  $.....$                                 & 2457163.20 $\pm$  3.89     & 2457293.95 $\pm$ 0.31  \\
$e$                       &  $.....$                                 & 0.368 $\pm$ 0.036          &  0.3272 $\pm$ 0.0095  \\ 
$\omega$ (deg)    &  $.....$                                & 236.90 $\pm$ 1.91               & 32.24 $\pm$ 1.05     \\
$\Omega$ (deg)   &  $.....$                                & 40.41 $\pm$ 2.82                 & 41.43 $\pm$ 1.09       \\
$a_1$ (mas)         &  $.....$                                & 0.377 $\pm$ 0.024           & 0.706 $\pm$ 0.014     \\
$i$ (deg)              &   $.....$                                & 102.25 $\pm$ 1.83               & 89.41 $\pm$ 0.96           \\
\hline                        
&& Other Parameters & \\
\hline                        

$D$ (pc)                   & 132.74 $\pm$ 0.12   & 133.33 $\pm$ 0.17      & 134.26 $\pm$ 0.17   \\
$m  ~(M_\odot)$\tablenotemark{b}  &  $.....$             & 2.9441             & 2.8745                \\
$m_1  ~(M_\odot)$  &  $.....$ & 2.8920 $\pm$ 0.0032  & 2.397 $\pm$ 0.010       \\
$m_2 ~(M_\odot)$   &  $.....$ & 0.0521 $\pm$ 0.0032 & 0.477 $\pm$ 0.010      \\
$a_1 ~(AU)$            &  $.....$ & 0.0503 $\pm$ 0.0032 & 0.0947 $\pm$ 0.0021  \\
$a_2 ~(AU)$           &  $.....$ & 2.7890 $\pm$ 0.0041 & 0.4760 $\pm$ 0.0019   \\

$\chi^2$, $\chi^2_{red}$   &  2309.29, 128.29 & 91.02, 8.27 & 70.80, 7.87  \\

%
\hline                                   
\end{tabular}
\tablenotetext{a}{ The parameters presented here were obtained with AGA. Very similar results were obtained with the least-squares fitting method. The sub-index 1 corresponds to the main component (i.e., the star) and the sub-index 2 corresponds to the secondary component (i.e., companion DoAr21$c$ or DoAr21$B$).}
\tablenotetext{b}{ The mass of the system is fixed and corresponds to the masses obtained from the Relative plus Absolute, simultaneous astrometry fit of components $b$ and $c$ (see Section 4.3). In the case of component $c$ we use the total mass obtained from the simultaneous fit of components $b$ and $c$. In the case of component $B$ we used the inner mass of the system, after removing the masses of components $b$ and $c$.}

\end{table*}



\begin{table*}[h]
\caption{Absolute Astrometry Fit: two companions\tablenotemark{a}}             
\label{tab:abs2fit}
\centering                          
\begin{tabular}{lc}        
\hline\hline                 
Parameter  & DoAr21$c$ and DoAr21$B$   \\    
\hline                        
& Fitted Parameters   \\
\hline                        

$\mu_{x}$ (mas/yr)  &  $-$19.6442  $\pm$   0.0035       \\
$\mu_{x}$ (mas/yr)  &  $-$26.8756  $\pm$   0.0084       \\

$\Pi$ (mas)              &   7.474          $\pm$   0.012         \\

$P$ (days)               &  1034.50        $\pm$   7.74           \\
$T_{0}$ (days)         &  2457157.64 $\pm$   6.94           \\
$e$                          &   0.404           $\pm$  0.065       \\
$\omega$ (deg)      &  240.43          $\pm$   3.52           \\
$\Omega$ (deg)     &  59.72            $\pm$   6.98           \\
$a_{1}$ (mas)         &    0.203          $\pm$   0.024        \\
$i$ (deg)                 &  109.99          $\pm$   5.62            \\

$P$ (days)               &   92.921         $\pm$   0.091           \\
$T_{0}$ (days)         &  2457295.80 $\pm$   0.94           \\
$e$                          &   0.391           $\pm$   0.029       \\
$\omega$ (deg)      &   39.61          $\pm$   2.78         \\
$\Omega$ (deg)     &  32.73            $\pm$   2.50          \\
$a_{1}$ (mas)         &   0.346          $\pm$   0.020        \\
$i$ (deg)                 &  93.03          $\pm$   2.03           \\
\hline                        
& Other Parameters  \\
\hline                        

$D$ (pc)                   & 133.80          $\pm$ 0.21          \\
$m  ~(M_\odot)$\tablenotemark{b}      &  2.9441        \\
$m_{1}  ~(M_\odot)$  &  2.681       $\pm$ 0.014     \\
$m_{2} ~(M_\odot)$   &  0.0279       $\pm$ 0.0031      \\
$m_{3} ~(M_\odot)$   &  0.236       $\pm$ 0.014     \\

$a_{1-2} ~(AU)$         &  2.869         $\pm$ 0.014    \\
$a_{1} ~(AU)$            &  0.0271       $\pm$ 0.0032     \\
$a_{2} ~(AU)$            &  2.842         $\pm$ 0.011      \\
$a_{1-3} ~(AU)$         &  0.5736       $\pm$ 0.0004     \\
$a_{1} ~(AU)$            &  0.0463       $\pm$ 0.0027     \\
$a_{3} ~(AU)$            &  0.5273       $\pm$ 0.0024       \\

$\chi^2$, $\chi^2_{red}$  &  31.72, 7.93    \\

%
\hline                                   
\end{tabular}
\tablenotetext{a}{ The parameters presented here were obtained with AGA. In this case, two components were fitted simultaneously (components $B$ and $c$) using the absolute astrometry model. The sub-index 1 corresponds to the main component (i.e., the star) and the sub-index 2 and 3 correspond to the companions (DoAr21$c$ and DoAr21$B$).}
\tablenotetext{b}{ The mass of the system is fixed and corresponds to the mass obtained from the Relative plus Absolute, simultaneous astrometry fit of components DoAr21$b$ and DoAr21$c$ (see Section 4.3).}

\end{table*}



\begin{table*}[h]
\caption{Relative plus Absolute Astrometry Fit\tablenotemark{a}}             
\label{tab:relfit}
\centering                          
\begin{tabular}{lc}        
\hline\hline                 
Parameter  & DoAr21$b$    \\    
\hline                        
& Fitted Parameters   \\
\hline                        

$\mu_{x}$ (mas/yr)  &  $-$19.6682 $\pm$   0.0030     \\
$\mu_{x}$ (mas/yr)  &  $-$26.9746 $\pm$   0.0074     \\

$acc_x$ (mas/yr/yr)  &       0.0057 $\pm$ 0.0021            \\ 
$acc_y$ (mas/yr/yr)  &  $-$0.0091 $\pm$ 0.0052           \\ 

$\Pi$ (mas)             &   7.350 $\pm$   0.010                \\
$P$ (days)              &  452.97 $\pm$   0.15                 \\
$T_{0}$ (days)        &  2457507.45 $\pm$  1.07           \\
$e$                          &   0.076 $\pm$    0.012              \\
$\omega$ (deg)       &  220.90 $\pm$   0.91                \\
$\Omega$ (deg)       &  53.20 $\pm$   1.78                 \\
$a_{1}$ (mas)          &    0.286 $\pm$   0.015              \\
$i$ (deg)                   &  103.57 $\pm$   1.17             \\
$q$   (m1/m2)           &   39.86 $\pm$    2.67              \\
$\omega_{2}$ (deg)  &  40.90 $\pm$   0.91                 \\
$a$ (mas)                  &  11.68 $\pm$   0.44                 \\
\hline                        
& Other Parameters  \\
\hline                        

$D$ (pc)               & 136.05 $\pm$ 0.19             \\
$m  ~(M_\odot)$  &  2.6064 $\pm$  0.0018    \\
$m_{1}  ~(M_\odot)$  &  2.5426 $\pm$ 0.0029     \\
$m_{2} ~(M_\odot)$  &  0.0638 $\pm$ 0.0035      \\
$a_{1-2} ~(AU)$  &  1.5886 $\pm$ 0.0007     \\
$a_{1} ~(AU)$  &  0.0389 $\pm$ 0.0021       \\
$a_{2} ~(AU)$  &  1.5497 $\pm$ 0.0014        \\

$\chi^2$, $\chi^2_{red}$  &  903.11, 64.51    \\

%
\hline                                   
\end{tabular}
\tablenotetext{a}{ The parameters presented here were obtained with AGA. Very similar results were obtained with the least-squares fitting method. The sub-index 1 corresponds to the main component (i.e., the star) and the sub-index 2 corresponds to the secondary component (i.e., companion DoAr21$b$).}

\end{table*}



\begin{table*}[h]
\caption{Relative plus Absolute, simultaneous Astrometry Fit\tablenotemark{a}}             
\label{tab:rel2fit}
\centering                          
\begin{tabular}{lcc}        
\hline\hline                 
Parameter  & DoAr21$b$ and $c$ & DoAr21$b$ and $B$   \\    
\hline                        
& Fitted Parameters  & \\
\hline                        

$\mu_{x}$ (mas/yr)  &  $-$19.6986  $\pm$   0.0036     &  $-$19.5478 $\pm$ 0.0036  \\
$\mu_{x}$ (mas/yr)  &  $-$26.9768  $\pm$   0.0088     & $-$26.8373 $\pm$   0.0088  \\

$acc_x$ (mas/yr/yr)  &       $.....$      &      0.0232 $\pm$ 0.0025      \\ 
$acc_y$ (mas/yr/yr)  &       $.....$     &  $-$0.0012 $\pm$ 0.0061     \\ 

$\Pi$ (mas)              &   7.428          $\pm$   0.012       & 7.385 $\pm$    0.012  \\

$P$ (days)               &  446.52         $\pm$   0.19         &  453.22 $\pm$   0.22  \\
$T_{0}$ (days)         &  2457045.94 $\pm$  1.22         & 2457063.16 $\pm$   1.38   \\
$e$                          &   0.090           $\pm$    0.036     & 0.095 $\pm$   0.018  \\
$\omega$ (deg)      &  250.02          $\pm$   1.11         & 225.86 $\pm$  1.20  \\
$\Omega$ (deg)     &  55.80            $\pm$   2.39         & 55.41 $\pm$  2.43  \\
$a_{1}$ (mas)         &    0.099          $\pm$   0.015       & 0.073 $\pm$  0.018  \\
$i$ (deg)                 &  104.19          $\pm$   1.52          & 105.43 $\pm$ 1.67  \\
$q$   (m1/m2)        &   121.91          $\pm$    19.16      & 151.89 $\pm$  37.59  \\
$\omega_{2}$ (deg)  &  70.02         $\pm$   1.11          & 45.86 $\pm$ 1.20   \\
$a_{1-2}$ (mas)        &  12.11            $\pm$   0.54           & 11.13 $\pm$  0.50   \\

$P$ (days)               &  984.93         $\pm$   6.39         & 92.935 $\pm$   0.041  \\
$T_{0}$ (days)         &  2457158.61 $\pm$   6.34         & 2457295.83 $\pm$   0.43   \\
$e$                          &   0.234           $\pm$   0.060      & 0.355 $\pm$   0.013  \\
$\omega$ (deg)      &  243.17          $\pm$   2.78         &  41.35 $\pm$  1.31  \\
$\Omega$ (deg)     &  44.12            $\pm$   4.36         &  46.06 $\pm$  1.60  \\
$a_{1}$ (mas)         &    0.321          $\pm$   0.026       & 0.650 $\pm$  0.018  \\
$i$ (deg)                 &  99.53          $\pm$   2.80          &  90.07 $\pm$ 1.34  \\
\hline                        
& Other Parameters & \\
\hline                        

$D$ (pc)                   & 134.62          $\pm$ 0.22        & 135.40    $\pm$  0.22     \\
$m  ~(M_\odot)$      &  2.9477         $\pm$  0.0025    &  2.2246  $\pm$  0.0021  \\
$m_{A}, m_{A}  ~(M_\odot)$  &  2.8782       $\pm$ 0.0027    &  1.838  $\pm$  0.018   \\
$m_{b}, m_{b} ~(M_\odot)$   &  0.0236       $\pm$ 0.0036    &  0.0146    $\pm$  0.0036    \\
$m_{c}, m_{B} ~(M_\odot)$   &  0.0459       $\pm$ 0.0032    &  0.372  $\pm$  0.012   \\

$a_{A-b}, a_{A-b} ~(AU)$         &  1.6308       $\pm$ 0.0009    & 1.5074 $\pm$  0.0010  \\
$a_{A}, a_{A} ~(AU)$            &  0.0133       $\pm$ 0.0020    & 0.0099 $\pm$  0.0024  \\
$a_{b}, a_{b} ~(AU)$            &  1.6175       $\pm$ 0.0011    & 1.4976 $\pm$  0.0014   \\
$a_{A-c}, a_{A-B} ~(AU)$         &  2.7778       $\pm$ 0.0092    & 0.3231 $\pm$  0.0004  \\
$a_{A}, a_{A} ~(AU)$            &  0.0433       $\pm$ 0.0036    & 0.0880 $\pm$  0.0025  \\
$a_{c}, a_{B} ~(AU)$            &  2.7346       $\pm$ 0.0056    & 0.4350 $\pm$  0.0021   \\

$\chi^2$, $\chi^2_{red}$  &  56.13, 6.24 & 60.32, 8.62   \\

%
\hline                                   
\end{tabular}
\tablenotetext{a}{ The parameters presented here were obtained with AGA. In this case, two components were fitted simultaneously (components $b$ and $c$, and components $b$ and $B$) using the combined model. The sub-index 1 corresponds to the main component (i.e., the star) and  sub-index 2 and 3 correspond to the two companions (i.e.,  DoAr21$b$ and DoAr21$c$, or DoAr21$b$ and DoAr21$B$).}

\end{table*}



\begin{table*}[h]
\caption{Weighted Average Parameters\tablenotemark{a} }             
\label{waverage}
\centering                          
\begin{tabular}{lccc}        
\hline\hline                 
Parameter & DoAr21$B$ & DoAr21$b$ & DoAr21$c$   \\    
\hline                        
&& Orbital Parameters  & \\
\hline                        
$P$ (days)            & 92.919 $\pm$ 0.022   & 450.88 $\pm$ 3.80      & 1013.5 $\pm$ 25.3  \\
$e$                       &  0.370 $\pm$ 0.035     & 0.089 $\pm$ 0.010     &  0.333 $\pm$ 0.090  \\ 
$\omega$ (deg)    &  38.55 $\pm$ 4.94      & 232.8 $\pm$ 15.6        & 240.54 $\pm$ 3.18     \\
$\Omega$ (deg)   &  38.67 $\pm$ 6.98     & 54.96 $\pm$ 1.41         & 51.1 $\pm$ 10.9       \\
$a_1$ (mas)         &  0.55 $\pm$ 0.20       & 0.15 $\pm$ 0.12           & 0.301 $\pm$ 0.026     \\
$i$ (deg)              &    91.31 $\pm$ 2.01    & 104.50 $\pm$ 0.96       & 105.75 $\pm$ 5.87           \\
\hline                        
&& Other Parameters & \\
\hline                        

$m_2 ~(M_\odot)$   &  0.35 $\pm$ 0.12 & 0.034 $\pm$ 0.026 & 0.042 $\pm$ 0.013      \\
$a_2 ~(AU)$           &  0.482 $\pm$ 0.046 & 1.550 $\pm$ 0.060 & 2.802 $\pm$ 0.056   \\

%
\hline                                   
\end{tabular}
\tablenotetext{a}{ The parameters presented are the weighted average values obtained from the fitted parameters presented in Table~\ref{tab:astrofit}  through Table~\ref{tab:rel2fit}. The estimated errors correspond to the standard deviation of these fitted values and reflects the dispersion from the different astrometric fits.
The sub-index 1 corresponds to the main component (i.e., the host star) and the sub-index 2 corresponds to the secondary component (i.e., companion DoAr21$B$, DoAr21$b$ or  DoAr21$c$).
}
\end{table*}



\begin{table*}[h]
\caption{Measured VLBA Source Fluxes}             
\label{fluxtab}
\centering                          
\begin{tabular}{rrr}        
\hline\hline                 
Julian Date & Peak Flux  & Flux Density    \\    
                  & (mJy)~~~         & (mJy)~~~    \\    

\hline                        
& Primary & \\
\hline                        
2453621.524 &  2.302 $\pm$ 0.254 &  5.939 $\pm$ 0.874 \\
2453744.188 &  0.425 $\pm$ 0.088 &  0.480 $\pm$ 0.166 \\
2453755.157 &  0.926 $\pm$ 0.090 &  1.003 $\pm$ 0.165 \\
2453822.972 &  1.215 $\pm$ 0.096 &  1.451 $\pm$ 0.187 \\
2453890.786 &  1.980 $\pm$ 0.104 &  2.930 $\pm$ 0.236 \\
2454092.235 &  2.541 $\pm$ 0.153 &  2.756 $\pm$ 0.281 \\
2454321.607 &  0.881 $\pm$ 0.077 &  1.874 $\pm$ 0.229 \\
2454331.079 &  1.724 $\pm$ 0.085 &  2.056 $\pm$ 0.166 \\
2454353.519 &  1.283 $\pm$ 0.077 &  1.756 $\pm$ 0.166 \\
2458257.821 &  4.402 $\pm$ 0.034 &  5.231 $\pm$ 0.067 \\
2458335.608 &  0.517 $\pm$ 0.025 &  0.681 $\pm$ 0.052 \\
2458534.064 &  4.388 $\pm$ 0.042 &  5.793 $\pm$ 0.088 \\
\hline                        
\hline                        
& Binary System & \\
\hline                        
& Primary & \\
\hline                        

2453691.332 &  6.075 $\pm$ 0.504 & 17.071 $\pm$ 1.85 \\
2453971.565 &  0.857 $\pm$ 0.090 &  1.137 $\pm$ 0.188 \\
2454365.487 &  2.499 $\pm$ 0.127 &  3.286 $\pm$ 0.265 \\
2458433.341 &  2.279 $\pm$ 0.172 &  5.355 $\pm$ 0.549 \\

\hline                        
& Secondary & \\
\hline                        

2453691.332 & 15.603 $\pm$ 0.510 & 35.654 $\pm$ 1.60 \\
2453971.565 &  1.084  $\pm$ 0.089 &  1.630  $\pm$ 0.203 \\
2454365.487 &  3.791  $\pm$ 0.129 &  4.107  $\pm$ 0.236 \\
2458433.341 & 24.311 $\pm$ 0.183 & 26.176 $\pm$ 0.333 \\

%
\hline                                   
\end{tabular}

\end{table*}


\section{Results}

\subsection{Single companion Astrometry}

\begin{figure*}
   \centering
   \plotone{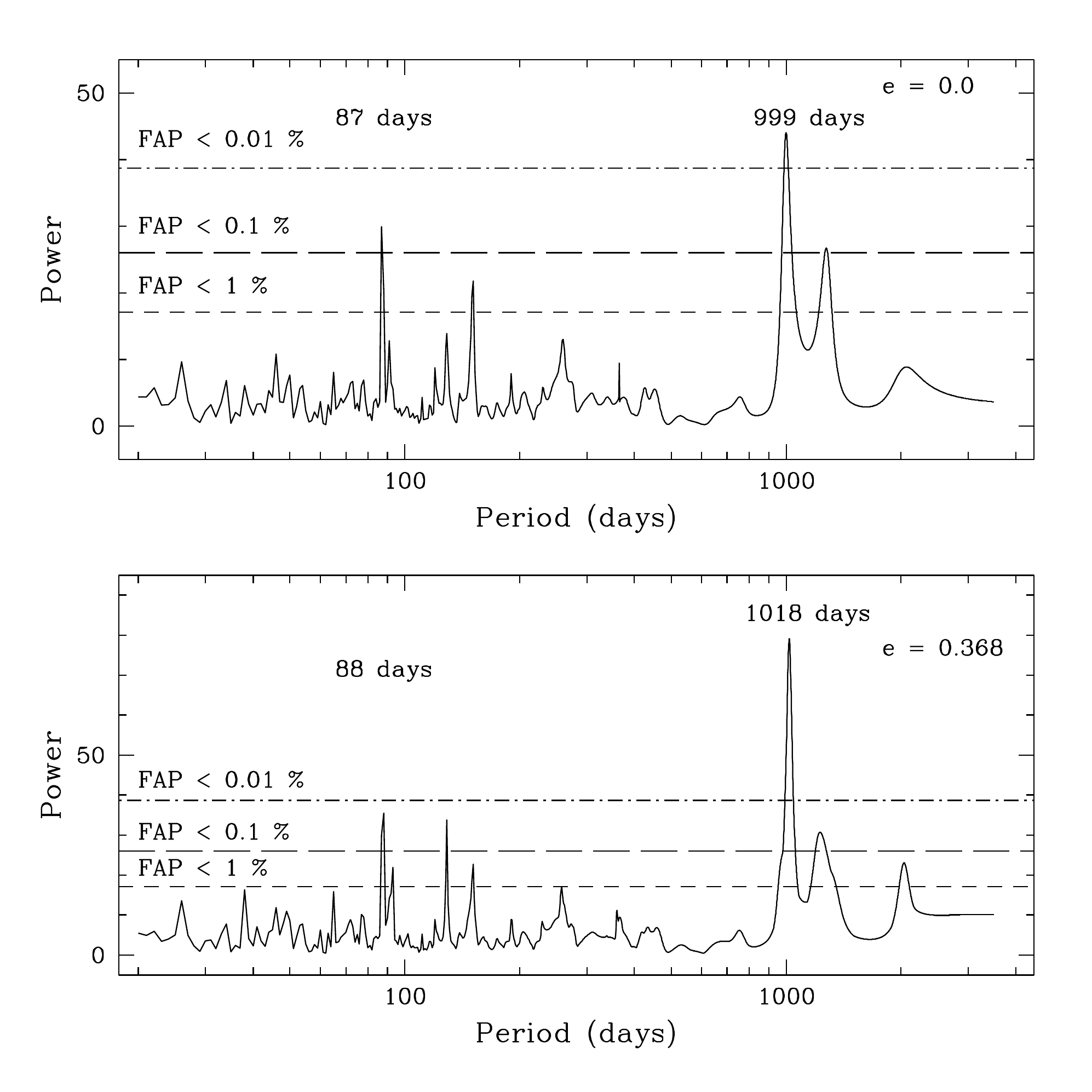}
    \caption{Least-Squares Periodogram. The upper panel shows the circular least-squares periodogram obtained by fixing the eccentricity e = 0. There is a prominent peak with a period of 999 days and a FAP of 0.004$\%$. There is an additional peak with a period of 87 days and a FAP close to 0.1$\%$. The lower  panel shows the periodogram that was obtained when fixing the eccentricity e = 0.368, which was obtained with the Absolute Astrometry fit of DoAr21$c$ (see Table~\ref{tab:astrofit}). 
The main peak becomes more prominent when including the eccentricity of the orbit. This indicates that this is a better fit to the data. The second peak also becomes marginally more prominent, suggesting that the orbit of this companion may also be eccentric.}
              \label{fig_2}%
    \end{figure*}

The least-squares periodogram with a circular orbit (CLSP) of the astrometric data (see Figure~\ref{fig_2}) shows a prominent peak (with 999 days period) with high power and very low False Alarm Probability (FAP). There is another significant peak (with 87 days period) with lower power and higher  FAP.
These is also a third weaker peak that appears in the CLSP periodogram (see Figure~\ref{fig_2}, upper panel) with a period of 151 days. 
However, when we fix the eccentricity to 0.368 (see discussion below), the peak with a period of about 999 days becomes more prominent,
while this weak peak remains constant and a somewhat stronger peak appears with a period of 129 days (see Figure~\ref{fig_2}). This new peak is quite weak in the upper panel of Figure~\ref{fig_2}. Thus these two weaker peaks are dubious, while the peak with a period of about 87 days remains consistent in both periodograms. 
Thus, we investigate here only the two candidates with the strongest peaks in the periodograms (with periods of about 999 and 87 days) to obtain their $''$significances$''$. 

\begin{figure*}
   \centering
    \plotone{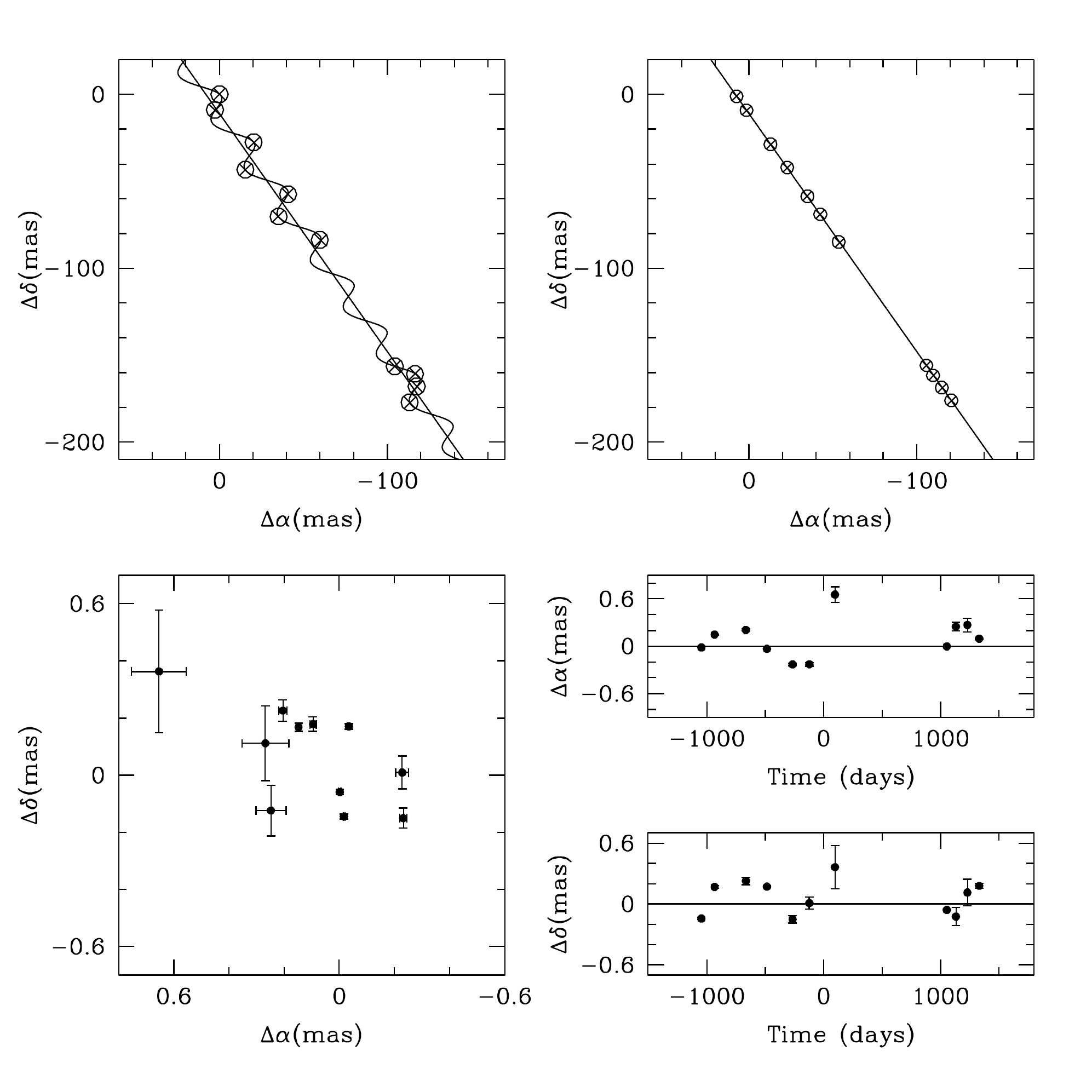}
    \caption{Parallax fit to the observed data. The upper$-$left panel shows the observed data and the astrometric fit obtained when fitting only the proper motions and the parallax of DoAr21. The upper$-$right panel shows the proper motions of the source after removing the parallax of the source. The lower$-$left panel show the residuals in RA and DEC, and the lower$-$right panels show the residuals as function of time. The residuals show a clear temporal trend that suggests that they could be due to at least one companion.}
              \label{fig_3}%
    \end{figure*}

We use the two methods described above (least-squares and AGA) to fit the astrometric data. First we use both methods to fit the 11 astrometric observations to obtain the proper motions and the parallax of  this source without taking into account any companion (Single Star Solution). The results of a single star solution are shown in Table~\ref{tab:astrofit} and Figure~\ref{fig_3}. 
We did not find necessary to include an acceleration term for the fitting of the external companion DoAr21$c$. However, in the case of the internal companion DoAr21$B$, we included acceleration terms to take into account the contribution of the external companion. The acceleration terms are small and produce a small effect in the fitting of DoAr21$B$.

\begin{figure*}
   \centering
   \plottwo{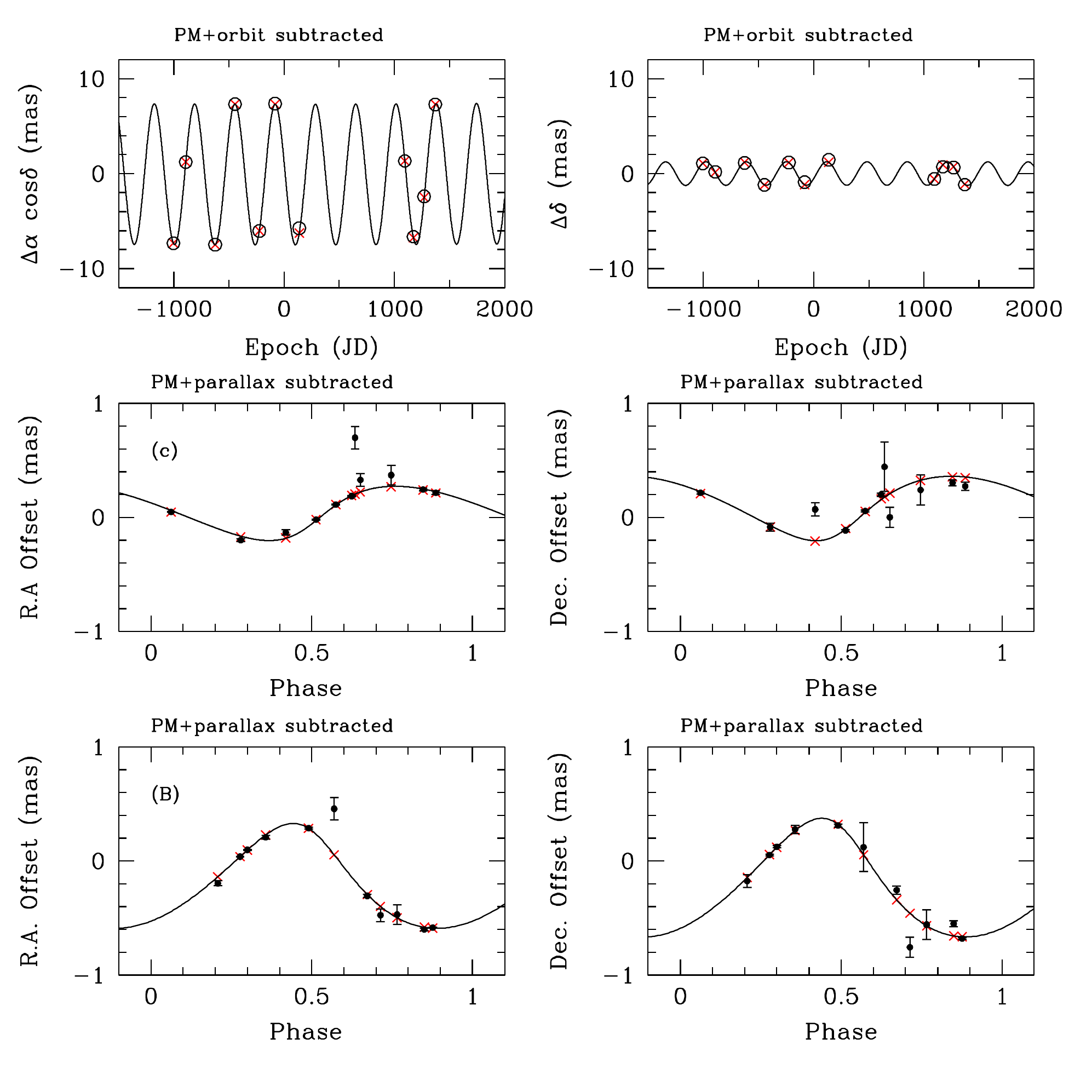}{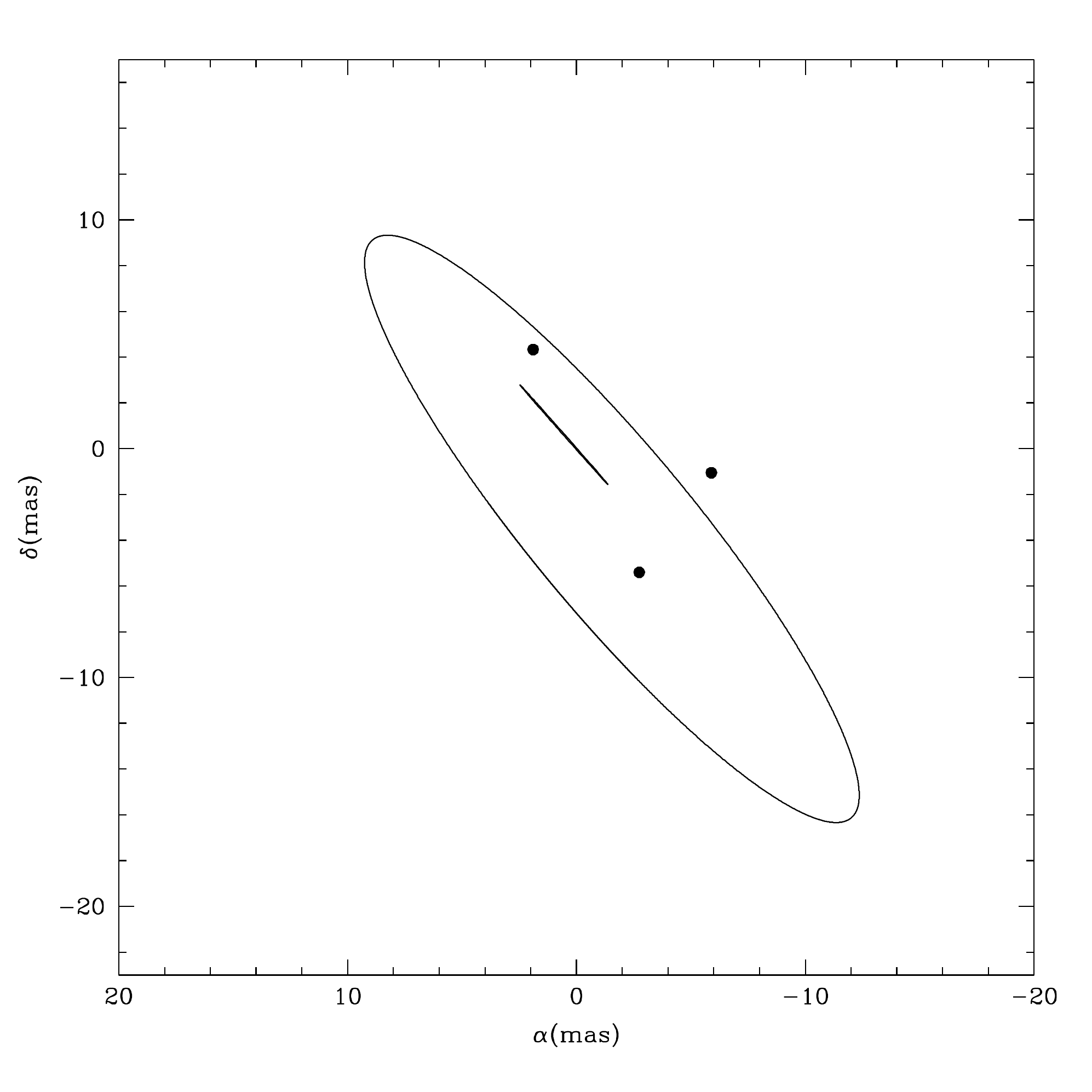}
   \caption{AGA fit of components DoAr21$c$ and DoAr21$B$. 
    LEFT. Top panels: Parallax of the source after removing proper motions and the solution of the orbit of DoAr21$c$. Next two panels: Orbital solution for DoAr21$c$ after removing proper motions and the parallax of the source. The last two panels show the orbital solutions for DoAr21$B$. The orbital solutions of the two companions are plotted as function of the orbital phase of each companion. The black dots correspond to the observed astrometric data with their error bars. The red crosses indicate the expected position at the observed epochs. The RA and DEC offsets are relative to the estimated barycenter position of the source. \\
    RIGHT: The solid lines show the estimated orbit of these two companions. This figure shows that the inclination angle of both companions is quite different. DoAr21$B$ has an inclination angle close to 90 degrees, while DoAr21$c$ has an inclination angle larger that 90 degrees. The black dots indicate the position of the companion detected with the VLBA. The error bars of these data points are smaller than the size of the dot. None  of the detections coincides with the estimated orbit of a companion.  
   }
              \label{fig_4}%
    \end{figure*}

We find that after the fitting the residuals are large compared with the noise present in the data and the astrometric precision obtained with the VLBA. 
Figure~\ref{fig_3} shows residuals up to about 0.3 mas and the astrometric precision obtained with the VLBA is about 30 $\mu$as. 
Furthermore, the residuals have a temporal trend that suggests the presence of at least one companion. We then use the least-squares and the AGA  algorithms to fit the astrometric observations of this source including a single companion (i.e., we performed an independent fit for each companion candidate). In both cases we obtain consistent solutions. The parallax and the orbital fits for the two candidates  are presented in Figure~\ref{fig_4}. Table~\ref{tab:astrofit} summarizes the best fits and their $\chi^{2}_{red}$ per degree of freedom 
($\chi^{2}_{red}$ = $\chi^{2}/(N_{data} - N_{par} - 1)$, where $N_{data} = 2 \times N_{points}$ and $N_{par}$ is the number of fitted parameters). 
The fit of the astrometric data clearly improves when including a companion, as seen by the $\chi^{2}_{red}$.  

The two components have similar  position angle of the line of nodes ($\Omega$), within the errors.
The eccentricity of the orbits seem to be reasonably well constrained: component $c$ has an orbital eccentricity $\sim$0.37, while component DoAr21$B$ has an eccentricity of $\sim$0.33.
The astrometric signature of the source ($a_{*}$) due to the gravitational pull of the companion is  larger in the case of component DoAr21$B$, by almost a factor of two. 
The inclination of the orbit ($i$) is somewhat different for the two companions. The inclination angle is larger than 90 degrees for component DoAr21$c$ ($\sim$102 deg, which indicates a retrograde orbit), while component $B$ has an orbital inclination of about 90 degrees (indicating an edge-on orbit), which suggests that the orbits of these two companions are not coplanar. This result is somewhat surprising since one would expect a similar inclination for the orbits of all the companions. We do not have an explanation for this result.
However, it is important to point out that the combined fit (relative astrometry plus absolute astrometry) of two components (components DoAr21$b$ and DoAr21$c$; see Section 4.3) also give retrograde orbits ($i$ $>$ 90 degrees).
With these fits we cannot estimate the dynamical mass of the system, thus to estimate the mass of the companions we use the mass of the system obtained with the combined fit of the astrometric data (a fixed mass  of $M$ $\sim$ 2.94 $M_\odot$ for the fit of component DoAr21$c$ and $\sim$ 2.87 $M_\odot$ for the fit of component DoAr21$B$). These masses were obtained with the combined fits of the orbits of components DoAr21$b$ and DoAr21$c$  (see Section 4.3). With this assumption, we obtain that the component with a compact orbit (component DoAr21$B$) has a mass consistent with a low mass star ($m_{B} \sim$ 0.48 $m_\odot$), while component DoAr21$c$ has a mass consistent with a Brown Dwarf ($m_{c} \sim$ 54.57 $m_{jup}$). The orbits of the two companions have semi-mayor axis $a_{c}$ $\sim$ 2.79 AU and $a_{B}$ $\sim$ 0.48 AU, respectively. This suggests that the mass of the companion decreases with the distance to the source. The Brown Dwarf is the component with the wider orbit.
The estimated distance to the source is  similar in both cases, however, the estimated distance is slightly larger when including a companion than when fitting only the proper motions and the parallax of the host star. 

The orbital periods of the orbits of components DoAr21$c$ and DoAr21$B$ are consistent with those found in the CLSP (see Figure~\ref{fig_2}). The difference in the orbital periods obtained with the astrometric fits and the CLSP is due to the fact that in the case of CLSP we assume circular orbits, while in the astrometric fits we include the eccentricity of the orbits as a free parameter to be fitted. In Figure~\ref{fig_2}, we also show the least-squares periodogram of the observed astrometric data fixing the eccentricity obtained for companion DoAr21$c$ ($e_{c}$ $\sim$ 0.368). In this case we obtain an orbital period for component DoAr21$c$ that is consistent with that obtained with the astrometry fit. This figure  shows that the power of component $c$ increases substantially when fixing the eccentricity of this companion. On the other hand, the power of component DoAr21$B$ does not change substantially when using this eccentricity in the periodogram.

Figure~\ref{fig_4} shows the orbits of both companions. As it was mentioned before, a companion was detected with the VLBA in 3 different epochs. The relative position of the companion with respect to the host star is included in Figure~\ref{fig_4}. 
This figure shows that the relative position of none of the detected companions coincide with the fitted orbits.
Bellow we attempt to fit simultaneously the orbits of the two inferred components (DoAr21$c$ and DoArt21$B$). We also attempt to fit simultaneously the relative astrometry of the companion detected with the VLBA (DoAr21$b$) and the absolute astrometry of the host star (combined fit; Section 4.3).

\subsection{Simultaneous Astrometric fit of DoAr21$c$ and DoAr21$B$}

\begin{figure*}
   \centering
   \plottwo{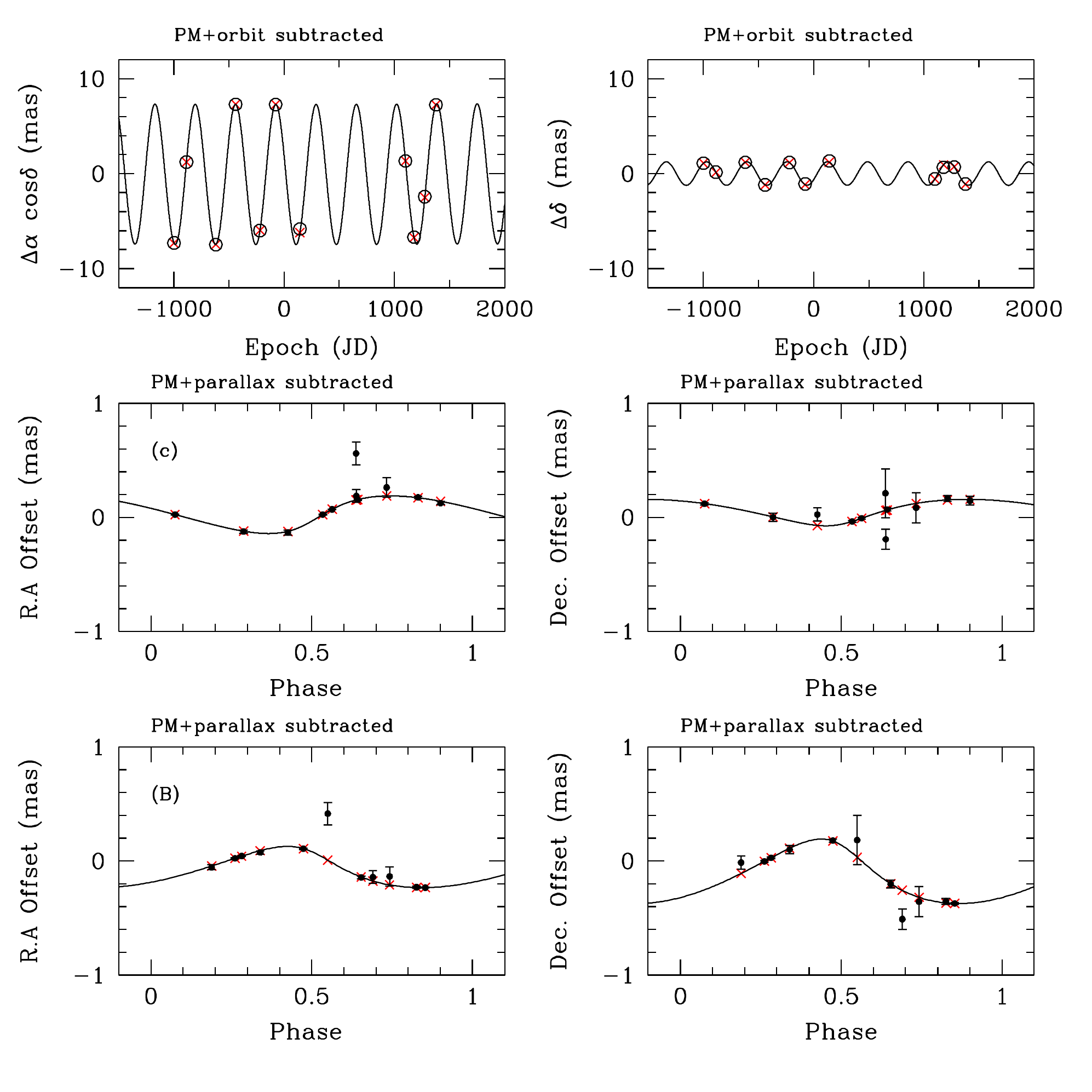}{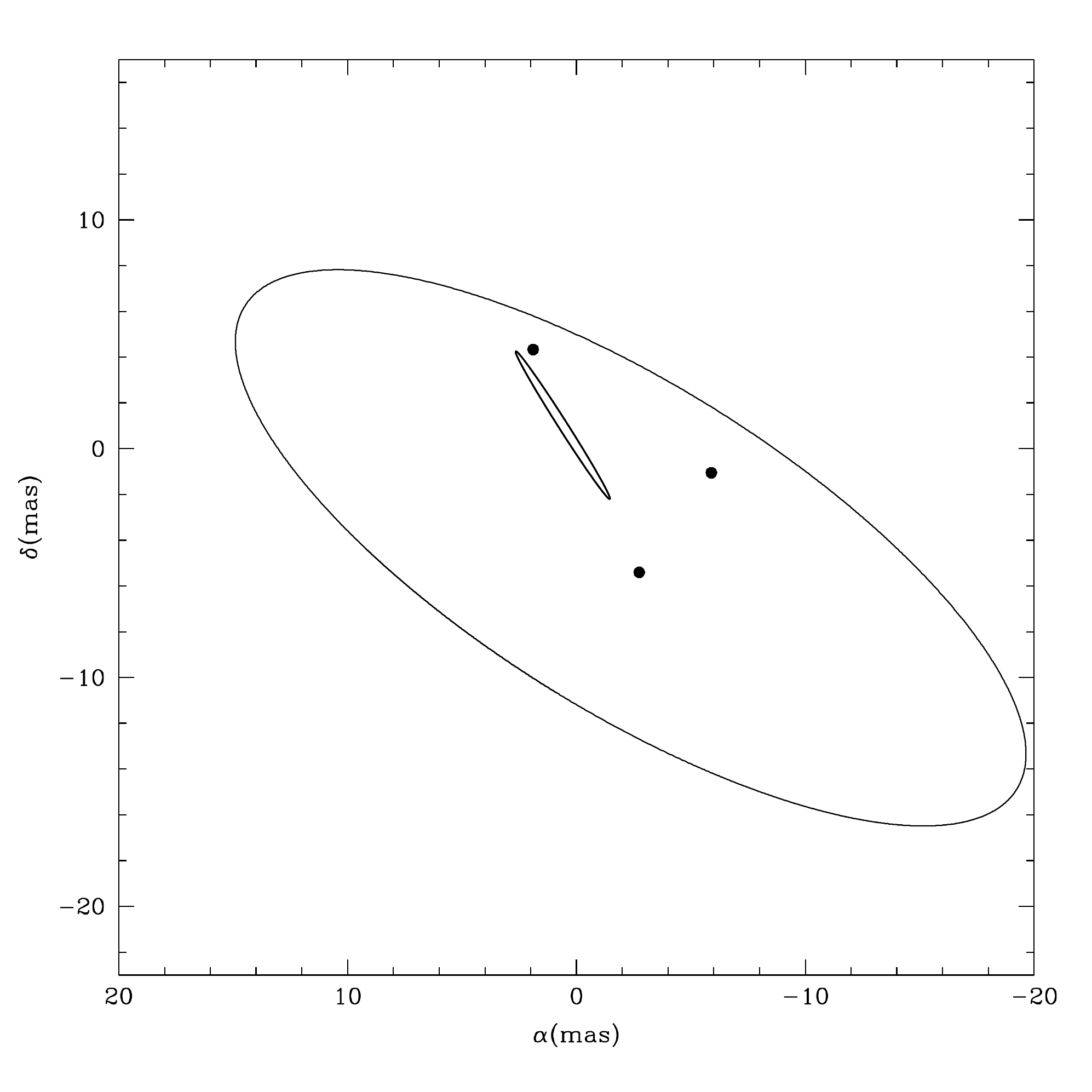}
   \caption{AGA simultaneous fit of components DoAr21$c$ and DoAr21$B$. 
   LEFT. Top panels: Parallax of the source after removing proper motions and the solution of the orbit of DoAr21$c$ and DoAr21$B$. Next two panels: Orbital solution for DoAr21$c$ after removing the proper motions and the parallax of the source, and removing the contribution of component DoAr21$B$. The last two panels show the orbital solutions for DoAr21$B$ after removing the parallax, the proper motions and the orbital motions of DoAr21$c$. The black dots correspond to the observed astrometric data with their error bars. The red crosses indicate the expected position at the observed epochs. The orbital solutions are shown as function of the orbital phase of each companion. The RA and DEC offsets are relative to the estimated barycenter position of the source. \\
   RIGHT. The solid lines show the estimated orbit of these two companions. This figure shows that the inclination angle of both companions is quite different. DoAr21$B$ has an inclination angle close to 90 degrees, while DoAr21$c$ has an inclination angle larger that 90 degrees. The black dots indicate the position of the companion detected with the VLBA. The error bars of these data points are smaller than the size of the dot. None  of the detections coincides with the estimated orbit of a companion. 
   }
              \label{fig_5}%
    \end{figure*}

The eleven observed epochs are in principle enough to fit simultaneously the orbits of the two companions (DoAr21$c$ and DoAr21$B$), plus the parallax and the proper motions of the host star. We find that the astrometry fit improves  when fitting simultaneously both companions. 
Table~\ref{tab:abs2fit} summarizes the parameters obtained with the best two-companion astrometry fit of the data. Figure~\ref{fig_5} shows the  parallax of the host star and the orbital motion of the host star due to the gravitational pull of both companions. This figure also shows the orbits of both companions.
For this fit we have also fixed the total mass of the system ($m = 2.9441$ $m_\odot$; see Section 4.3 and Table~\ref{tab:rel2fit}). The fitted parameters are similar to those found from the single-companion fit. However, some of the fitted parameters change substantially. For instance, the eccentricity, semi-mayor axis and inclination of the orbits are somewhat larger, and the estimated masses of the companions are somewhat smaller in this case than in the case of a single-companion astrometry fit. Furthermore, the position angle of the line of nodes of both companions differs by more than 20 degrees ($\Omega$ $\sim$ 32.7 and 59.7 for DoAr21$B$ and DoArt21$c$, respectively), while in the case of a single-companion fit we obtained a similar $\Omega$ for both companions ($\Omega$ $\simeq$ 41 deg). These new results suggest that the orbits of these two companions are not coplanar.

In addition, Figure~\ref{fig_5} shows that the companion detected with the VLBA does not coincide with the estimated orbits of these two companions. We obtained a similar result from the single-companion astrometry fit. This suggests that there is probably another companion that does not appear in the periodogram, but that was detected with the VLBA.

Bellow we attempt to fit simultaneously the relative astrometry of the companion observed with the VLBA  and the absolute astrometry of the host star (combined fit).

\subsection{Relative plus absolute Astrometry: Combined$-$fit}

\begin{figure*}
   \centering
    \plottwo{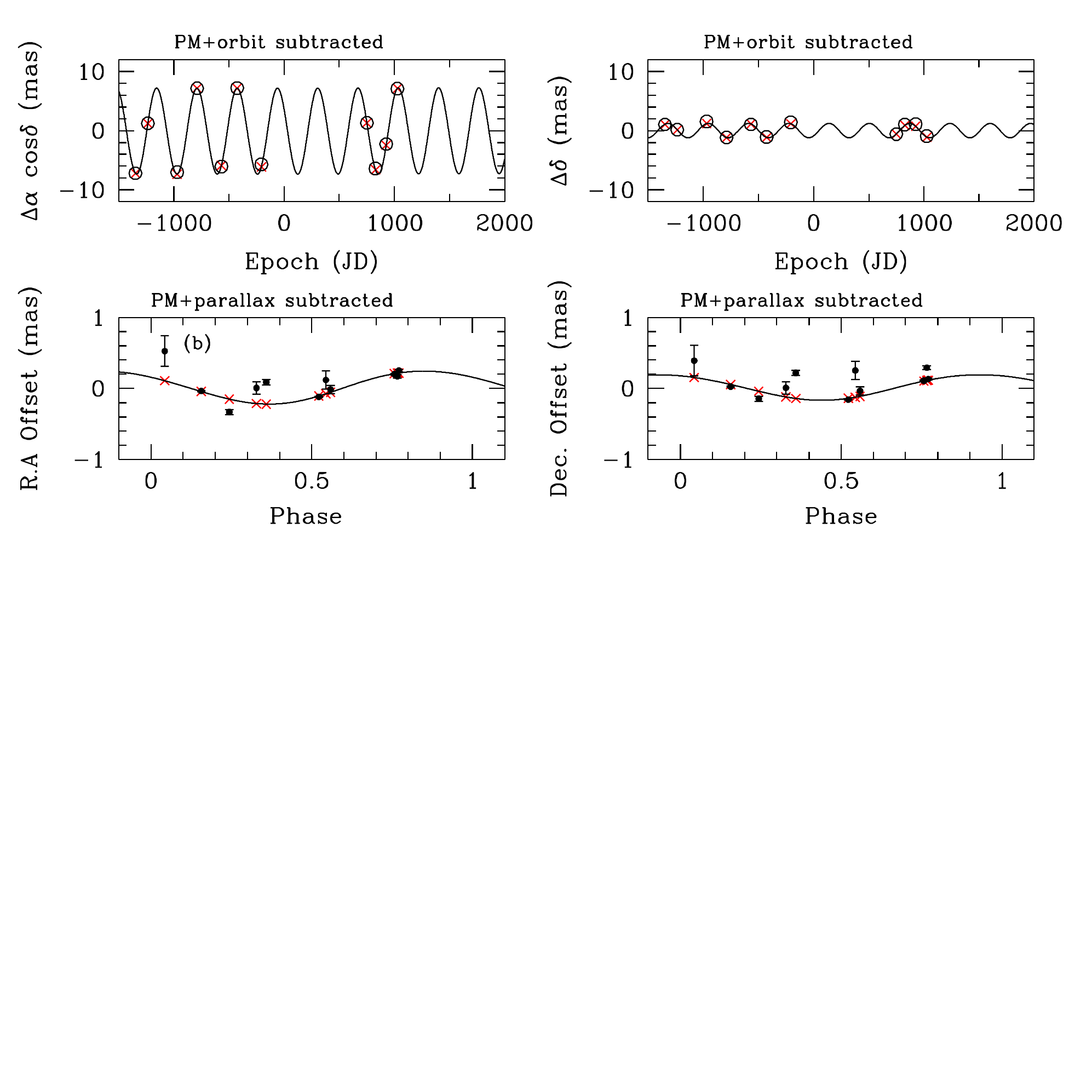}{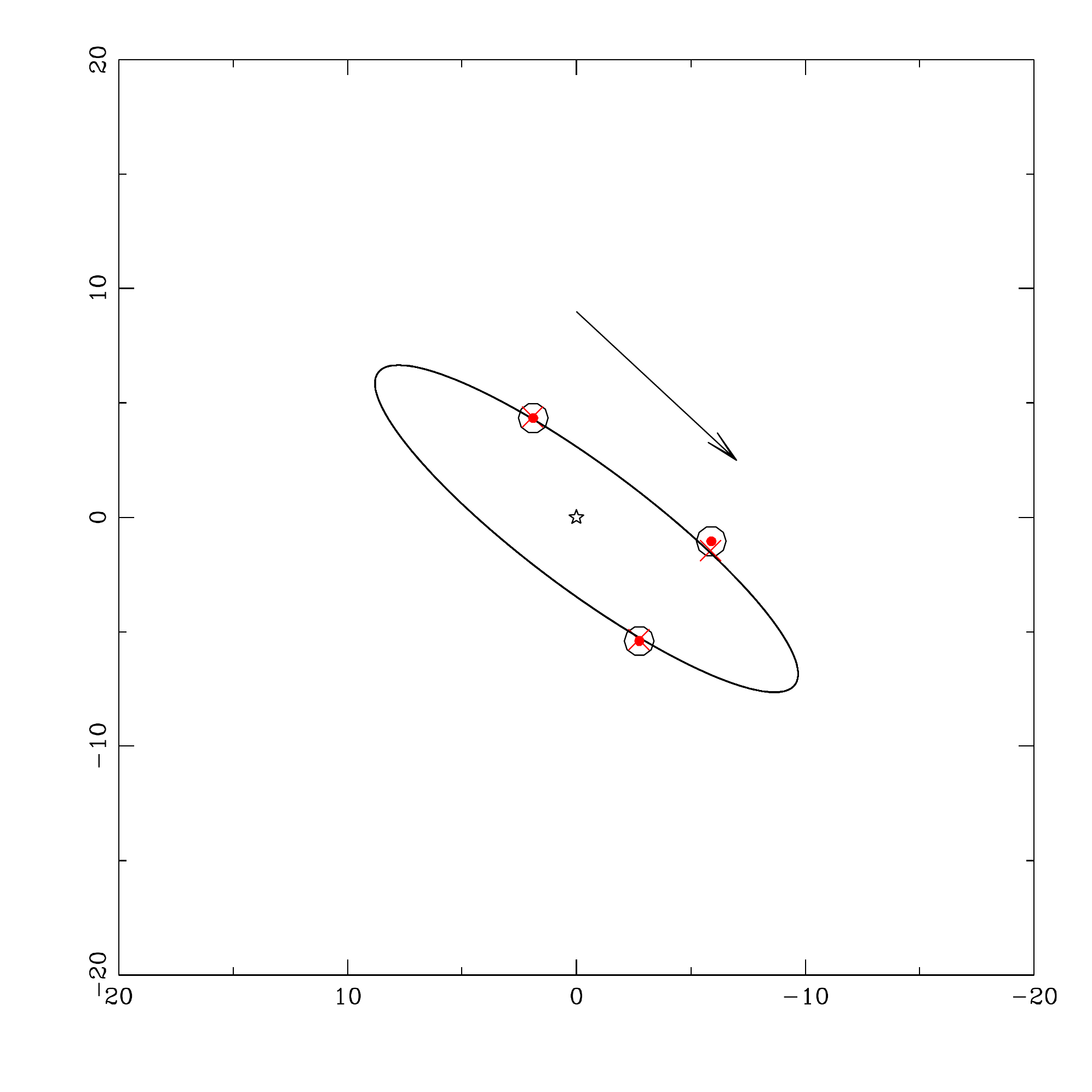}
    \caption{AGA fit of  DoAr21$b$ obtained with the combined model. 
    LEFT. The upper panels show the parallax after removing proper motions and the contribution of the orbit of DoAr21$b$.  The lower  panels show  the orbit of  DoAr21$b$ after removing proper motions and parallax. The black dots correspond to the observed astrometric data with their error bars. The red crosses indicate the expected position at the observed epochs. The orbit of this companion is plotted as function of the phase orbit. The RA and DEC offsets are relative to the estimated barycenter position of the source. \\
    RIGHT. The solid line shows the orbit of  DoAr21$b$ around the host star. The open circles indicate the estimated positions of  DoAr21$b$  with respect to the host star (black star at the center of the orbit), and the red crosses indicate the predicted positions along the orbit at the observed epochs. The red dots indicate the positions of the observed companion. The error bars of these data points are smaller than the size of the dot.  There is an excellent agreement between the observed positions and the predicted positions.
    }
              \label{fig_6}%
    \end{figure*}

We investigate here the possibility that the relative positions of the detected companion are associated to another companion,  different from the two companions that we have already found.
We find that the three secondary detections with the VLBA can be fitted by combining relative astrometry and absolute astrometry (combined model) using both the least-squares and the AGA algorithms. These three detections correspond to  a new component that we call  DoAr21$b$. 
Table~\ref{tab:relfit} summarizes the parameters that were obtained with the best fit of the data, and the $\chi^{2}_{red}$ per degree of freedom.
The parallax and the astrometric signature of the host star due to the gravitational pull of this companion is  presented in Figure~\ref{fig_6}. The orbital fit of this companion is also presented in Figure~\ref{fig_6}.
Since DoAr21$b$ is an internal companion, we also included acceleration terms to take into account the contribution of the external companion DoAr21$c$. The acceleration terms are small and produce a small effect in the fitting of DoAr21$b$.

The combined fitted orbit of  DoAr21$b$  has a period of $P_{b}$ $\sim$ 453 $days$, a semi-mayor axis of $a_{1}$ $\sim$ 0.29 $mas$ (or $\sim$ 0.04 $AU$), a position angle of the line of nodes $\Omega$ $\sim$ 53 deg, an eccentricity $e$ $\sim$ 0.08, an inclination angle $i$ $\sim$ 104 $deg$, and a longitude of the periastron $\omega$ $\sim$ 221 deg. Component DoAr21$b$ has an orbit with a semi-mayor axis of $a_{b}$ $\sim$ 11.40 $mas$ (or $\sim$ 1.55 $AU$) and  a longitude of the periastron $\omega_{b}$ $\sim$ 41 deg. Since we are doing a combined fit (relative plus absolute orbital fit), we obtain the orbital fit around the barycenter position of the system for both, the main source and the companion, thus we obtain the dynamical mass of the system, as well as the masses of each individual component. The dynamical mass of the system is $M_{\star b}$ $\sim$ 2.606 $M_\odot$,  the mass of the main source is $M_{\star}$ $\sim$ 2.542 $M_\odot$, and the mass of the companion is $M_{b}$ $\sim$ 0.064 $M_\odot$ (or about 66.8 $M_{jup}$). 
The orbit of this new companion, DoAr21$b$, lies between the orbits of the other two companions, and its estimated mass is consistent with being a Brown Dwarf.

We find that the position angle of the line of nodes ($\Omega$) and the inclination angle ($i$) of the orbit are similar to those found in DoAr21$c$ obtained with the two-companion simultaneous fit (see Table~\ref{tab:abs2fit}). The eccentricity  of the orbit ($e$) is smaller than those found for the other companions, and seems to be well constrained. The astrometric signature of the source ($a_{\star}$) due to the gravitational pull of this companion is somewhat smaller than those found for the other companions (see Table~\ref{tab:astrofit},  \ref{tab:abs2fit} and \ref{tab:relfit}). Since the astrometric signature of the source is relatively small, the residuals of the fit (see Figure~\ref{fig_6}) are larger than those obtained from the fits of the other two companions (see Figure~\ref{fig_5}). 
In addition, since the astrometric signature of DoAr21$b$ is quite small ($<$ 0.1 $mas$; see Tables~\ref{tab:relfit} and \ref{tab:rel2fit}) the least-squares fit of the orbit of this companion basically fails. In other words, the $\chi^{2}$ of the fit is similar to the null solution, and thus the estimated power is small, within the expected $''$noise$''$ in the periodogram (see discussion in section 3.1).
These results are consistent with the fact that companion DoAr21$b$ does not appear in the periodogram (see Figure~\ref{fig_2}). In other words, we have found DoAr21$b$ only because this companion was detected at several epochs with the VLBA, otherwise the astrometric signature due to this companion would be embedded in the residuals of the fits of the other two companions.

\begin{figure*}
   \centering
    \plottwo{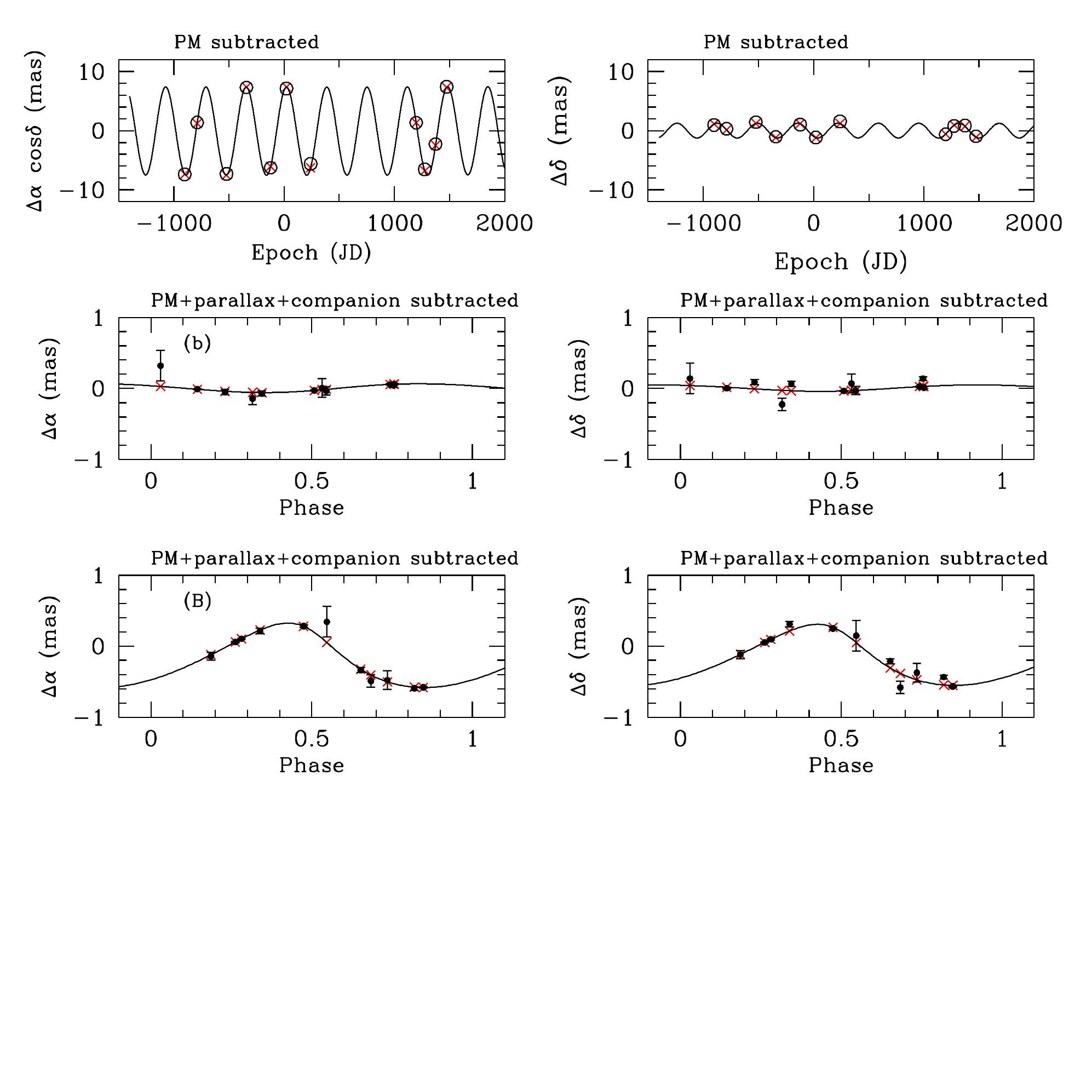}{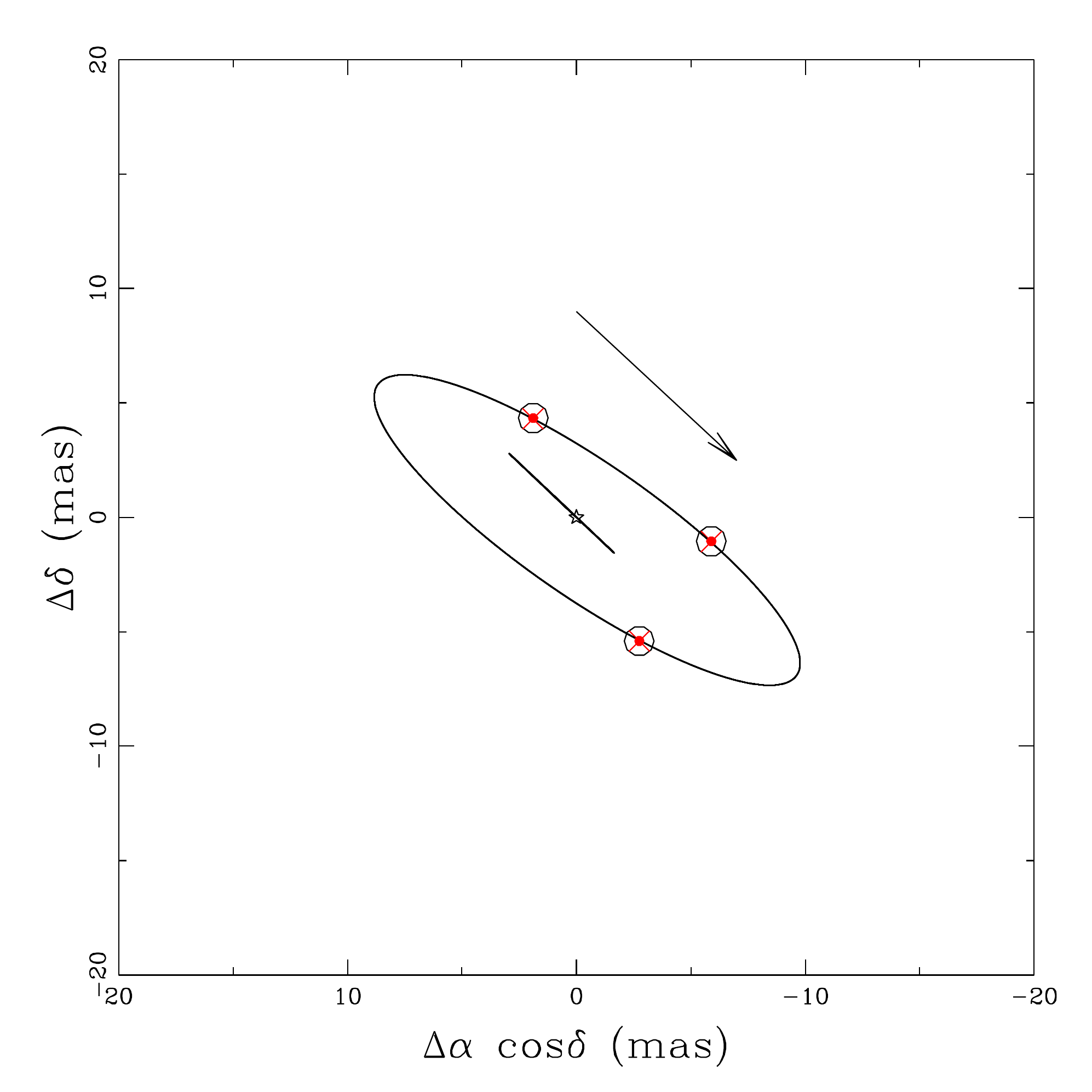}
    \caption{Simultaneous AGA fit of components DoAr21$b$ and DoAr21$B$ using the combined model, and fitting simultaneously both components (se Section 4.3). 
    LEFT. The upper panels show the parallax after removing proper motions and the orbital contribution of both components. 
The next two panels show  the orbit of DoAr21$b$ after removing proper motions, parallax and the contribution of DoAr21$B$. The lower panels show the orbit of  DoAr21$B$ after removing proper motions, parallax and the contribution of DoAr21$b$. The black dots correspond to the observed astrometric data with their error bars. The red crosses indicate the expected position at the observed epochs. The RA and DEC offsets are relative to the estimated barycenter position of the source. \\
    RIGHT. The solid lines show the orbits of  components DoAr21$b$ and DoAr21$B$ around the host star. The open circles indicate the estimated positions of  DoAr21$b$ with respect to the host star (black star at the center of the orbits), and the red crosses indicate the predicted positions along the orbit at the observed epochs. The red dots indicate the positions of the observed companion. The error bars of this data points are smaller than the size of the dot. In the case of DoAr21$b$, there is an excellent agreement between the observed positions and the predicted positions. The arrow indicates the direction of the movement of DoAr21$b$ along its orbit. Note that the orbit of DoAr21$B$ is eccentric and perpendicular to the plane of the sky.
    }
              \label{fig_8}%
    \end{figure*}

The eleven observed epochs and the three direct detections of DoAr21$b$ are in principle enough to fit simultaneously the orbits of two companions (DoAr21$b$ and DoAr21$c$, or DoAr21$b$ and DoAr21$B$), plus the parallax and the proper motions of the host star. 
Table~\ref{tab:rel2fit} summarizes the parameters that were obtained with the best fit of the data, and the $\chi^{2}_{red}$ per degree of freedom.
The parallax and the astrometric signature of the host star due to the gravitational pull of each companion is  presented in Figures~\ref{fig_7} and \ref{fig_8}. The orbital fit for each pair of companions is also presented in these Figures. 
For the simultaneous fit of DoAr21$b$ and DoAr21$B$, we included acceleration terms to take into account the contribution of the external companion DoAr21$c$. The acceleration terms are small and produce a small effect in the fitting of these two companions.
We find that the fitted parameters are in general consistent with the previous fits. In particular, the astrometric fit of  DoAr21$b$ is quite consistent in all the fits of this component. This is because the relative astrometry of this companion, using the three detected epochs, constrain the orbit of this companion. The astrometric fits of the other two companions are similar to those obtained previously, however, there are some differences. For instance, the position angle of the line of nodes ($\Omega$), the inclination angle ($i$) and the semi-mayor axis ($a_{1}$) of the orbit are somewhat different from those found previously (see Table~\ref{tab:abs2fit}). It is important to mention that the estimated mass of the system is somewhat different in both fits, even if we take into account the mass of the companion DoAr21$c$ in the simultaneous fit of DoAr21$b$ and DoAr21$B$. The difference in the estimated total mass is $\sim$ 0.72 $M_\odot$ (or $\sim$ 0.68 $M_\odot$, taking into account the mass of DoAr21$c$ in both fits), which is much higher than the estimated mass of DoAr21$c$ ($\sim$ 0.046 $M_\odot$). A better estimate of the masses in this multiple system can be obtained by fitting simultaneously the orbits of all the components in the system. Thus, further observations of this multiple system will be needed in order to obtain a better estimate of the total mass of the system and the individual masses of all the companions.  

\section{Discusion}

\subsection{Mass and spatial distribution in this multiple system}

\begin{figure*}
   \centering
    \plottwo{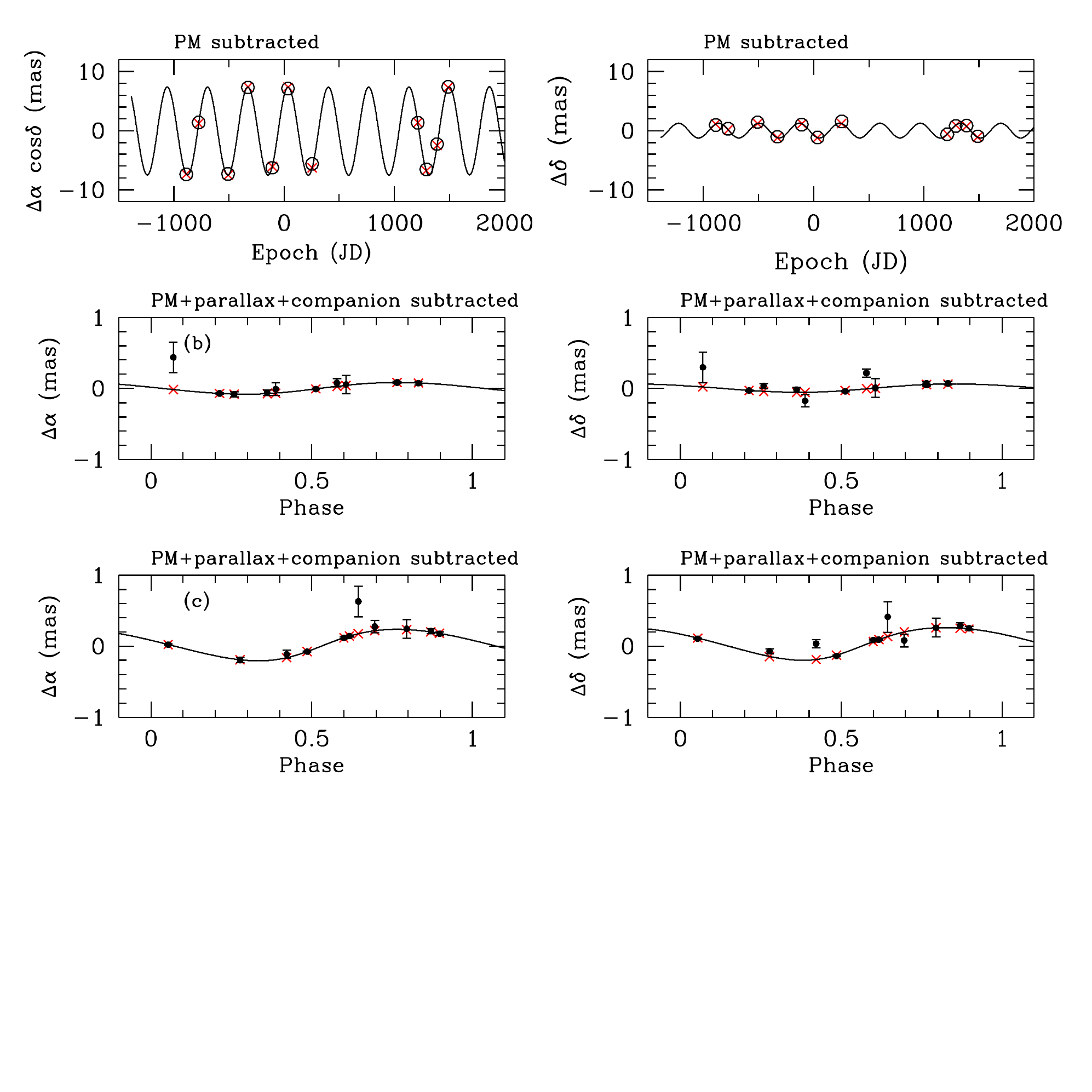}{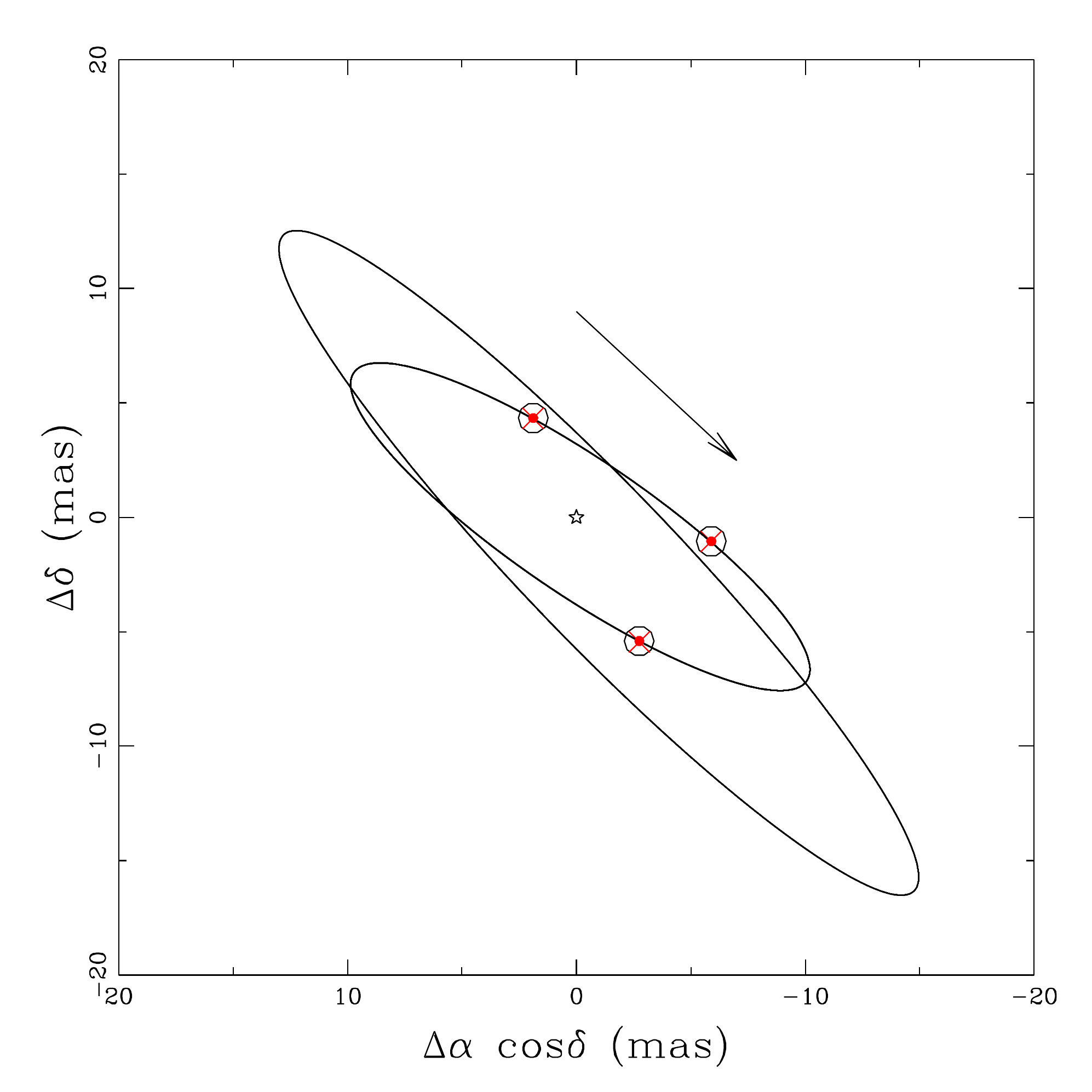}

    \caption{Simultaneous AGA fit of components DoAr21$b$ and DoAr21$c$ using the combined model,  and fitting simultaneously both components (see Section 4.3). 
    LEFT. The upper panels show the parallax after removing proper motions and the orbital contribution of components DoAr21$b$ and DoAr21$c$. 
The next two panels show  the orbit of the DoAr21$b$ after removing proper motions, parallax and the contribution of DoAr21$c$. The lower panels show the orbit of  DoAr21$c$ after removing proper motions, parallax and the contribution of DoAr21$b$. The black dots correspond to the observed astrometric data with their error bars. The red crosses indicate the expected position at the observed epochs. The RA and DEC offsets are relative to the estimated barycenter position of the source. \\
    RIGHT. The solid lines show the orbits of  components DoAr21$b$ and DoAr21$c$ around the host star. The open circles indicate the estimated positions of  DoAr21$b$ with respect to the host star (black star at the center of the orbits), and the red crosses indicate the predicted positions along the orbit at the observed epochs. The red dots indicate the position of the observed companion. The error bars of this data points are smaller than the size of the dot. There is an excellent agreement between the observed positions and the predicted positions. The overlap of the two orbits is a projection effect due to the large inclination of the orbits and the difference in the line of nodes ($\Omega$) of each orbit (see Table~\ref{tab:rel2fit}). The arrow indicates the direction of the movement of components DoAr21$b$ and DoAr21$c$ along their orbits.
    }
              \label{fig_7}%
    \end{figure*}

Figure~\ref{fig_7} shows the astrometric signature of the host star due to the companions DoAr21$b$ and DoAr21$c$, and the parallax of the system as function of time, and Figure~\ref{fig_8} shows the astrometry of DoAr21$b$ and DoAr21$B$. 
Figure~\ref{fig_7} suggests that the orbits of DoAr21$b$ and DoAr21$c$ cross each other, however, the results presented in Table~\ref{tab:rel2fit} indicate that this is a projection effect due to the large inclination of the orbits ($i$ $\sim$104 and $\sim$100 $deg$, respectively) and the difference in the position angle of the line of nodes of both orbits ($\Omega$ $\sim$56 and $\sim$44 $deg$, respectively). 
The semi-mayor axis of the orbit of DoAr21$c$ ($a_{c}$ $\sim$2.73 $AU$) is nearly a factor of two larger than that of DoAr21$b$ ($a_{b}$ $\sim$1.62 $AU$), and their orbits have low eccentricities ($e_{b}$ $\sim$0.09 and  $e_{c}$ $\sim$0.23).
Thus the orbits of these two companions do not cross each other.
In addition, the orbit of DoAr21$B$ ($a_{B}$ $\sim$0.44 $AU$) is about a factor of three smaller than that of DoAr21$b$ ($a_{b}$ $\sim$1.50 $AU$). Therefore the orbits of this multiple system seem to scale with a relationship close to 3:1 between DoAr21$B$ and DoAr21$b$, and close to 2:1 between DoAr21$b$ and DoAr21$c$. 

The inclination angle of the three companions is not the same. The orbit of DoAr21$B$ appears to be  edge-on ($i$$\sim$90.1 deg), while the orbits of DoAr21$b$ and DoAr21$c$ are retrograde, with inclination angles $\sim$104 and 100 deg, respectively. 
Thus the inclination angle of the system seem to be about 97$\pm$7 deg. However, this large difference in the inclination angle will have to be confirmed with further observations.

Since a companion (DoAr21$b$) was detected in three epochs, we were able to fit simultaneously the relative position of the companion and the absolute position of the host star. Furthermore, we were able to fit simultaneously the orbits of the pairs of companions DoAr21$b$$-$DoAr21$c$ and DoAr21$b$$-$DoAr21$B$.
With this relative plus absolute simultaneous combined fit, we have obtained the dynamical mass of the system, as well as the mass of the individual companions. 
The combined fit of the pairs of components give the dynamical masses for the system of $\sim$2.948 $M_\odot$ for the pair DoAr21$b$$-$DoAr21$c$ and $\sim$2.225 $M_\odot$ for the pair DoAr21$b$$-$DoAr21$B$. 
There is a significative difference in the estimated mass of the system of about 0.68 $M_\odot$ (taking into account the mass of DoAr21$c$ in both fits). 
This discrepancy is probably due to the astrometric contribution of the other companion (which is not taken into account in the fit), and the fact that DoAr21$b$ was detected at only three epochs.
We consider that at a first approximation of the total mass of the system is probably somewhere between these two values. 

There is also a large difference in the estimates mass of the host star, between 1.84 and 2.51 $M_\odot$ (we have taken into account the mass of DoAr21$B$ in both estimates; see Table~\ref{tab:rel2fit}). The difference of about 0.67 $M_\odot$ is similar to the difference of about 0.68 $M_\odot$ found for the total mass of the system. Thus, we also consider that the mass of the star is somewhere between these two estimated masses. The estimated mass of DoAr21 is similar to the previously estimated masses of $\sim$$2.2 \,M_\odot$ and $\sim$$1.8 \,M_\odot$ \citep{jensen09}, which were obtained by assuming that all the luminosity of the source is associated to a single star and by splitting the luminosity of this source in two equal stars, respectively. The  dynamical mass that we obtained here is, at present, the best estimated mass of the WTTS DoAr21.

Tables~\ref{tab:astrofit}  through  \ref{tab:rel2fit}  show that the best fitted parameters of the three companions of DoAr21 do not change considerably when including one or more companions in the astrometric fit.
However, taking into account that the different fits give slightly different values for all the orbital parameters, and that the orbits of the companions are not yet fully constrained, we have calculated the weighted average of the orbital parameters for all the companions, as well as the weighted average values for the mass of the host star and the companions. The weighted average parameters are presented in Tables~\ref{waverage}.

The estimated masses of the host star and the three companions are: 2.04 $\pm$ 0.70 $M_\odot$ for the host star,  0.35 $\pm$ 0.12 $M_\odot$ for DoAr21$B$, 0.034 $\pm$ 0.026 $M_\odot$ (35.6 $\pm$ 27.2 $M_{jup}$) for DoAr21$b$ and 0.042$\pm$ 0.013 $M_\odot$ (44.0 $\pm$ 13.6  $M_{jup}$) for DoAr21$c$ (see Table~\ref{waverage}). The inner companion, DoAr21$B$, has an estimated mass consistent with a low mass star,  while the other two companions have masses consistent with being Brown Dwarfs. 

The masses of DoAr21$b$ and DoAr21$c$ are probably consistent with a spectral types M7-M8 (e.g., \citet{luhman09}, and references therein), however, the classification of Brown Dwarfs is based on their spectral type, in particular based on the elements seen in their optical spectrum. 
These two Brown Dwarfs can not be classified this way since  they, in principle, can not be separated from the host star in the optical nor in the infrared. In addition, it is expected that  brown Dwarfs change in class type as they burn their deuterium (and their lithium in the case of the most massive Brown Dwarfs). 
Since DoAr21 is estimated to have an age os about 0.4$-$0.8 Myrs \citep{jensen09}, the two Brown Dwarfs orbiting this WTTS must be extremely young. 
In addition, these two Brown Dwarfs have broad orbits, with semi-mayor axis of $\sim$1.6 and $\sim$2.8 $AU$ (see Table~\ref{waverage}).
These results suggest that these Brown Dwarfs were formed far away from the host star, probably close to their current orbits. 
This is consistent with the idea that these Brown Dwarfs were formed by fragmentation of the disk where this young star was formed. 
On the other hand, it is not clear if DoAr21$B$, being the most massive companion ($\sim$ 0.35 $M_\odot$) in the system and with the most compact orbit ($\sim$ 0.482 $AU$), was also formed by fragmentation of the circumstellar disk, neither if it was formed close to its current orbit.
 
\subsection{Variability in this multiple system}

\begin{figure*}
   \centering
   \plotone{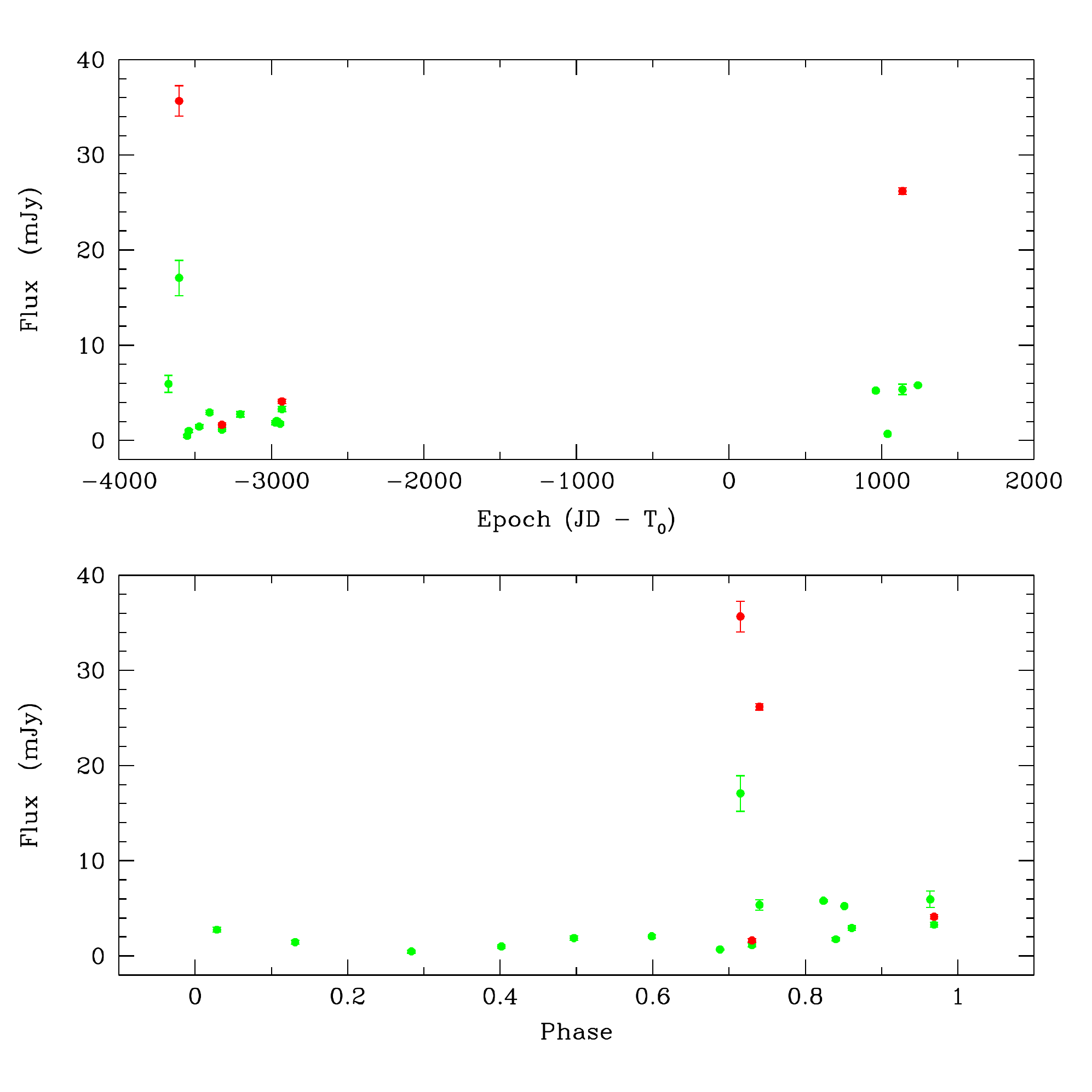}
    \caption{Time variability of DoAr21 at radio wavelengths (see Table~\ref{fluxtab}). Upper panel shows the flux density variation as function of time. The flux density of the primary source (the host star) is in green and the flux density of the secondary source (the detected companion) is in red. This figure shows flux variations of several mJy (between a fraction of a mJy and up to about 6 mJy) in the time span of the observations of more than 4500 days. DoAr21 had two strong outburst with more than 20 mJy. $T_0$ = 2457295.83 days corresponds to the time of the passage through the periastron of the orbit of DoAr21$B$ (see Table~\ref{tab:rel2fit}).
 Lower panel shows the flux variation of DoAr21 as function of the orbital phase of the low-mass stellar companion DoAr21$B$. A phase equal to 0.5 corresponds to the time at which DoAr21$B$ passes the periastron of its orbit.}
    
              \label{fig_9}%
    \end{figure*}

DoAr21 presents a large variability at radio wavelengths (see Table~\ref{fluxtab}).  Figure~\ref{fig_9} shows that the radio flux of this WTTS changes in more than two orders of magnitude, from a fraction of a mJy to several tens of mJy. This figure shows that there are abrupt changes in the flux in intervals of several weeks. However, this change could occur in much shorter times, the cadence of the observations is not adequate to find how steep are the outbursts observed in this source, neither if there is a frequency in the outbursts.
It has been speculated that this variability is probably due to the interaction between the magnetosphere of the star an the magnetosphere of a close stellar companion. We have found three astrometric companions to DoAr21. 
DoAr21$B$ is the most massive companion ($m_{B}$$\sim$0.35 $M_\odot$) and with the most compact and eccentric orbit ($a_{B}$ $\sim$0.482 $AU$, $e_{B}$$\sim$0.37) in this multiple system.  
Thus, DoAr21$B$ could be the responsible for the large flux variation observed in DoArt21. 
Given the large eccentricity of the orbit of this companion, and its proximity to the star, the interaction of their magnetospheres would probably occurs when the companion is closer to the star, which occurs close to the periastron of the orbit. 

To investigate this possibility, we have first calculated the closest distance between DoAr21$A$ and DoAr21$B$, which occurs close to the periastron of their orbital motions. We obtain that the closest distance between them is $a_{p} = a(1 - e) = 0.34$ $AU$, where $a = a_{\star} + a_{B}$ $\sim$ 0.536 $AU$ is the relative semi-major axis of the orbit of DoAr21$B$ around the host star DoAr21, and $e$ $\sim$ 0.37 is the eccentricity of the orbit (see Table~\ref{waverage}). The closest separations between the two stars is much larger than the typical magnetosphere sizes of the stars, which are in the range of a few stellar radii (e.g.,  \citet{feigelson99}). 
Second, we investigated the posibility of a correlation between the period of the orbit of DoAr21$B$ and the variability of the star. We have calculated the expected position of DoAr21$B$ in its orbit for each observed epoch. 
In Figure~\ref{fig_9}, we present the observed fluxes of DoAr21 folded with the period of DoAr21$B$ ($P_{B}$$\sim$93.0 days). 
We do not find a clear correlation between the outbursts observed in DoAr21 and the orbital period of this component. Figure~\ref{fig_9} shows that the two strongest peak fluxes nearly coincide in the orbital phase of this companion, about halfway between the periastron and the apoastron in the orbit of DoAr21$B$. 
However, one would expect that the outbursts would occur when the low-mass stellar companion is closest to the star (near the periastron of its orbit). If DoAr21$B$ were the responsable of the outbursts, then the outburst occurs about 20 days after this low-mass star has passed its periastron. 
These results indicate that DoAr21$B$ is probably not the responsable for the variability neither the X-ray outburst observed in DoAr21. This suggests that there may be another companion with a more compact orbit than those of the companions we report here, that perturbes the magnetosphere of the host star.
Further observations will be needed to search for this putative companion in a very compact orbit around DoAr21. 

The sub-stellar companion DoAr21$b$ was detected at three epochs at radio wavelengths with the VLBA. This suggests that this Brown Dwarf is highly variable at radio wavelenghts. The other Brown Dwarf in this system, DoAr21$c$, which is in a more extended orbit than DoAr21$b$, was not detected at any of the observed epochs. It is not clear why DoAr21$b$ was directly detected with the VLBA, while DoAr21$c$ was not.

This is the first time that a Brown Dwarf orbiting a young star has been directly detected.
Several Brown Dwarfs  have been detected with the VLA and with the VLBA, but most of them are single stars (they are not part of an stellar system) and  are located close by (a few tens of parsecs away), and have ages of more that 100 Myrs (e.g., \citet{berger02}, \citet{sahlmann13} and \citet{forbrich16}). 
A few Brown Dwarf candidates have been found at radio wavelengths (e.g., \citet{rodriguez17}, \citet{dzib13}), and they are also isolated sources. 
A putative Brown Dwarf was recently found to be orbiting a TTauri star \citep{ginski18}, however its orbit lyes outside the circumbinary disk, with a projected orbit of about 210 $AU$. The estimated mass of this companion is quite uncertain, and it could be a very low-mass star.

\subsection{Distance to DoAr21}

Tables~\ref{tab:astrofit}  through  \ref{tab:rel2fit}  show that the estimated distance to DoAr21 does not change substantially when including one or more companions in the astrometric fit.
Taking into account that the different fits give slightly different values for the distance, and that the orbits of the companions are not yet fully constrained, we have calculated the weighted average of the estimated distances. We obtained that the weighted distance is $d = 134.6 \pm 1.0$ ~pc, 
where the estimated error corresponds to the standard deviation of the fitted values, which better reflects the dispersion seen in the different astrometric fits.
This estimate is an improvement to the distance to this source of 135.76 $\pm$2.27 pc, previously obtained with the VLBA \citep{ortizleon18}. 
The distance to DoAr21 that we obtain is in agreement, within the estimated errors,  with that recently obtained  by \citet{ortizleon17}.
The error that we obtain here, however, it is a few times smaller than those obtained previously for this source, and the typical errors for the other sources in the Ophiucus Complex, obtained by \citet{ortizleon17}.  This is probably due to the larger number of observations used for the present astrometric fit, the accuracy of the observations we present here (see Table~\ref{tab:jmfit}) and a better coverage of the parallax, as well as the inclusion of the astrometric signal of one or more companions, which was not taken into account by \citet{ortizleon17}.

\subsection{Expected Radial Velocities}

We have obtained the astrometric best fits for the independent Keplerian orbits of the three DoAr21 companions. These solutions can be used to estimate an expected induced radial velocity of the star due to the gravitational pull of each companion. Assuming a simple model of totally independent companions (e.g., \citet{canto09}), the expected induced radial velocity would be:

\begin{equation}
K_{j} = \left(\frac{2 \pi G}{T_{j}}\right)^{1/3} \frac{m_{j} sin(i_{j})}{(M_{\star} + m_{j})^{2/3}}   \frac{1}{\sqrt{1 - e^{2}_{j}}},
\end{equation}

\noindent
where $G$ is the Gravitational constant, and $T_{j}$, $M_{\star}$, $m_{j}$ and $e_{j}$ are the orbital period, the star and companion masses and the eccentricity of the orbit of the companion. Using the parameters presented in sections 5.1 and 5.3 and those presented in Table 7, we obtain that the maximum induced velocities on DoAr21 by DoAr21$B$, DoAr21$b$ and DoAr21$c$ are  $K$ $\sim$ 9.902, 0.564 and 0.557 km s$^{-1}$, respectively. These radial velocities could in principle be observed with high-spectral resolution spectrographs. However, TTauri stars are magnetically active, may have  broad (molecular) spectral features, and present ubiquitous variability, that would make very difficult to observe such radial velocities. Recent optical spectroscopic observations of DoAr21 have shown possible velocity variations of $\sim$4.9  km s$^{-1}$ over the course of about 2 hours, but with a quite low precision ($\sim$1-2 km s$^{-1}$) due to the high rotation velocity of this star \citep{james16}. This velocity variation, if real, may suggest a much shorter orbital period than that estimated for DoAr21$B$. 
Future short and long term, high-resolution spectroscopic observations of DoAr21 may show whether this short period signal is real or not, and furthermore, may be able to detect the $\sim$9.9 km s$^{-1}$ radial velocity signature that we find that DoAr21$B$ probably induces on DoAr21.

\section{Conclusions}

The multi-epoch VLBA observations of DoAr21 that we present here allow us to carry out a precise analysis of the spatial wondering of this source due to its parallax, proper motions and the astrometric signature of several companions. The precise astrometric observations obtained with the VLBA are crucial to carry out this kind of study. We find that the determination of the distance to this source improves significantly when the orbital motions of its multiple companions are taken into account. We also find that DoAr21 is a highly variable radio source, with continuos small variations in its flux (within a few mJy), and episodic outbursts (of a few tens of mJy).

We present here different ways to analyze the VLBA astrometric observations of the WTTS DoAr21. We have searched for possible companions using a least-squares periodogram and analysing the residuals that result from the astrometric fit of the proper motions and the parallax of the multi-epoch data of this source. We have used two different algorithms (a least-squares algorithm and a Genetic algorithm) to fit the astrometric multi-epoch data obtained with the VLBA, and to obtain the parallax  and proper motions of the host star, and the parameters of the orbits of possible companions. 
The periodogram of the astrometric data shows two astrometric signatures. The strongest  signature corresponds to the sub-stellar companion DoAr21$c$ and the weaker signature corresponds to the low-mass stellar companion DoAr21$B$. A third companion, DoAr21$b$, was directly detected with the VLBA.
We find that the best astrometric fits of the data are those where we fit simultaneously more than one companion (those that appear in the periodogram), and those were we combine relative and  absolute astrometric fits of the data, using simultaneously the relative astrometry of DoAr21$b$ and the absolute astrometry of the host star.
However, the multi-companion fit is limited by the reduced number of astrometric data of this source. Further VLBA observations of this source will allow to simultaneously fit the orbits of all the detected companion of the source.

Since DoAr21$b$ was directly detected with the VLBA in several epochs, we were able to obtain an accurate determination of the astrometric mass of the system, as well as  the masses of the individual components (the star and the companions).
We find that the WTTS DoAr21 is a multiple system, formed by a compact binary system (the host star and a low-mass star), and two sub-stellar companions, whose masses are consistent with being Brown Dwarfs and located in external orbits, all of them within 3 $AU$ from the host star.

Since this WTTS is very young (less than a million years old), we speculate that the companions of this source must also be extremely young, and that they were probably formed close to their present orbit.

%

%

%

%

%

%

%

%

%

%

\begin{acknowledgements}
We thank the referee for his/her valuable comments that helped to improve this paper. We thank Sergio A. Dzib for a detailed reading of an early version of the manuscript and valuable suggestions.
S.C. acknowledges support from DGAPA grant IN103318, UNAM and CONACyT, M\'exico.
G.N.O.-L. acknowledges support from the von Humboldt Stiftung.
The Long Baseline Observatory is a facility of the National Science Foundation operated under cooperative agreement by Associated Universities, Inc. The National Radio Astronomy Observatory is a facility of the National Science Foundation operated under cooperative agreement by Associated Universities, Inc.
\end{acknowledgements}


\bibliographystyle{aasjournal}


{}


\end{document}